\documentclass{aa}
\usepackage{graphicx}
\usepackage{natbib}
\bibpunct{(}{)}{;}{a}{}{,} 
\usepackage{txfonts}
\begin{document}
   \title{The puzzling dredge-up pattern in NGC\,1978}

   \author{M.~T.~Lederer
          \inst{1}
          \and
           T.~Lebzelter
          \inst{1}
          \and
          S.~Cristallo
          \inst{2}
          \and
          O.~Straniero
          \inst{2}
          \and
          K.~H.~Hinkle
          \inst{3}
          \and
          B.~Aringer
          \inst{1}
          }

   \offprints{M. T. Lederer}

      \institute{University of Vienna, Department of Astronomy,
              T\"urkenschanzstrasse 17, A-1180 Vienna, Austria\\
              \email{lederer@astro.univie.ac.at,lebzelter@astro.univie.ac.at,aringer@astro.univie.ac.at}
                            \and
                            INAF, Osservatorio Astronomico di Collurania, 64100 Teramo, Italy\\
                     \email{cristallo@oa-teramo.inaf.it,straniero@oa-teramo.inaf.it}
                     \and
                     National Optical Astronomy Observatories\thanks{Operated by the Association of Universities for
                     Research in Astronomy, under cooperative agreement with the
                     National Science Foundation.}, PO Box 26732, Tucson, AZ 85726, USA\\
                     \email{hinkle@noao.edu}
             }

   \date{Received February 17, 2009; accepted June 2, 2009}


  \abstract
   {Low-mass stars are element factories that efficiently release their products in the final stages of their evolution by means of stellar winds. Since they are large in number, they contribute significantly to the cosmic matter cycle. To assess this contribution quantitatively, it is crucial to obtain a detailed picture of the stellar interior, particularly with regard to nucleosynthesis and mixing mechanisms.
   }
   {We seek to benchmark stellar evolutionary models of low-mass stars. In particular, we measure the surface abundance of \element[][12]{C} in thermally pulsing AGB stars with well-known mass and metallicity, which can be used to infer information about the onset and efficiency of the third dredge-up.
   }
   {We recorded high-resolution near-infrared spectra of AGB stars in the LMC cluster NGC\,1978. The sample comprised both oxygen-rich and carbon-rich stars, and is well-constrained in terms of 
the stellar mass, metallicity, and age.
We derived the C/O and \element[][12]{C}/\element[][13]{C} ratio from the target spectra by a comparison to synthetic spectra. Then, we compared the outcomes of stellar evolutionary models with our measurements.
   }
   {The M stars in NGC\,1978 show values of C/O and \element[][12]{C}/\element[][13]{C} that can best be explained with moderate extra-mixing on the RGB coupled to a moderate oxygen enhancement in the chemical composition. These oxygen-rich stars do not seem to have undergone third dredge-up episodes (yet). The C stars show carbon-to-oxygen and carbon isotopic ratios consistent with the occurrence of the third dredge-up. We did not find S stars in this cluster. None of the theoretical schemes that we considered was able to reproduce the observations appropriately. Instead, we discuss some non-standard scenarios to explain the puzzling abundance pattern in NGC\,1978.
   }
   {}


   \authorrunning{Lederer et al.}
   \titlerunning{The puzzling dredge-up pattern in NGC\,1978}
   \maketitle
%

\section{Introduction}\label{sect:introduction}
Modelling the final phases in the evolution of a low-mass star is
a demanding task. The stellar interior is a site of rich
nucleosynthesis, particularly when the star evolves on the
asymptotic giant branch (AGB). The freshly synthesised elements
are carried to the outer layers of the star by means of the {\em
third dredge-up} \citep[TDU, for a review
see][]{1999ARA&A..37..239B}. The onset and efficiency of this
mixing mechanism depends on the mass and the metallicity of the
star. Various stellar evolution models developed in the recent past
\cite[e.\,g.][]{1997ApJ...478..332S,2000A&A...360..952H,2004MNRAS.352..984S,2006NuPhA.777..311S,2007PASA...24..103K}
agree on the theoretical picture of the TDU, but details are still
subject to discussions. It is thus important to test the models
against observations and to derive constraints  
to improve our understanding of the involved phenomena and related
problems. AGB stars are tracers of intermediate-age stellar
populations, which can only be modelled accurately when the
evolution of the constituents is known. Moreover, evolved low-mass
stars undergo strong mass loss in the late stages of their
evolution. They enrich the interstellar medium with the products
of nuclear burning and thus, as they are large in number, play a
significant role in the cosmic matter cycle.

The primary indicator of TDU is a carbon surface
enhancement. An originally oxygen-rich star can be transformed
into a carbon star by an efficient dredge-up of 
\element[][12]{C} to the surface. The criterion to distinguish
between an oxygen-rich and a carbon-rich chemistry is the number
ratio of carbon to oxygen atoms, $\mbox{C/O}$. During the
thermally pulsing (TP) AGB phase, while the $\mbox{C/O}$ ratio is 
constantly rising as a consequence of the TDU,
there are mechanisms that may counteract
the increase in the carbon isotopic ratio
\element[][12]{C}/\element[][13]{C}. The radiative gap between the
convective envelope and the hydrogen burning shell can be bridged
by a slow mixing mechanism
\citep[cf.][]{1995ApJ...447L..37W,2003ApJ...582.1036N}. In this
way, \element[][12]{C} and isotopes of other elements are fed back
to nuclear processing. The physical origin of this
phenomenon is still not known, which is another argument to
establish observational data in order to benchmark theoretical
predictions.

Measurements of the carbon abundance and the carbon isotopic ratio
have been done for a few bright field AGB stars
\citep{1986ApJS...62..373L,1987ApJ...316..294H,1990ApJS...72..387S}.
The direct comparison of the values inferred from field star
observations to evolutionary models is complicated by the
fact that luminosity and mass of those targets can be determined
only inaccurately, while both quantities are crucial input
parameters for the models. A strategy to circumvent this problem
is to observe AGB stars in globular clusters (GC). They provide
a rather homogeneous sample with respect to distance, age, mass,
and metallicity. Admittedly, this simplistic picture is slowly
disintegrating. There is evidence that globular clusters often
harbour more than one population with abundance variations among
the constituents
\citep{2007MNRAS.379..151M,2007ApJ...661L..53P,2008MNRAS.391..354R}.
The parameters like mass and metallicity are, however, still much
better constrained than for field stars. For our purpose, i.\,e.
investigating the abundance variations due to the TDU, the old
Milky Way globular clusters are not well suited. The AGB stars of the
current generation have an envelope mass too low for TDU to occur,
while more massive stars have evolved beyond the AGB phase. The
intermediate-age globular clusters in the Large Magellanic Cloud
(LMC) serve our needs better. We started with an investigation of
the AGB stars in NGC\,1846 \citep[][henceforth
paper~I]{2008A&A...486..511L} and indeed demonstrated the
abundance variation along the AGB due to the TDU.

In this paper, we pursue this idea and study the AGB stars in the
LMC globular cluster NGC\,1978, deriving $\mbox{C/O}$ and
\element[][12]{C}/\element[][13]{C} ratios. We give an overview of
previous studies of this cluster and our observations in
Sect.~\ref{sect:observations}. Details about the data analysis are
given in Sect.~\ref{sect:dataanalysis}. In
Sect.~\ref{sect:evolutionarymodels} we describe the
evolutionary models we use for a comparison to our
observational results. The results follow in
Sect.~\ref{sect:results}. We discuss our findings in
Sect.~\ref{sect:discussion}, giving scenarios on how to interpret
the puzzling case of NGC\,1978 before we conclude in
Sect.~\ref{sect:conclusions}.


\section{Observations}\label{sect:observations}

\subsection{The target cluster: NGC\,1978}\label{sect:targetcluster}
The globular cluster NGC\,1978 is a member of the intermediate-age
($\tau=1-3\,\mathrm{Gyr}$) cluster population
\citep{1995A&A...298...87G} in the LMC. This population
corresponds to one of the two peaks in the LMC metallicity
distribution at $\mathrm{[Fe/H]}=-0.37$ (with a spread of
$\sigma=0.15$) reported by \citet{2005AJ....129.1465C}. In a
recent work, \citet{2007AJ....133.2053M} indeed derived a
metallicity of $\mathrm{[M/H]}=-0.37$ that is based on the
findings from \citet{2006ApJ...645L..33F}, who claimed
$\mathrm{[Fe/H]}=-0.38\pm0.02$ and that $\mathrm{[\alpha/H]}$ is
almost solar. In contrast to that, \citet{2000A&A...364L..19H}
quoted a grossly deviating value of $\mathrm{[Fe/H]}=-0.96\pm0.20$.
They investigated two stars from this cluster (LE8 and LE9, see
also Sects.~\ref{sect:dataanalysis} and \ref{sect:results}) and 
concluded that $\mathrm{[O/Fe]}=+0.37\pm0.10$.
These findings have not been reproduced by other groups.
\citet{2008AJ....136..375M} rather found a mild depletion in alpha
elements on average and ascribe the differences to
\citet{2000A&A...364L..19H} to the deviating metallicity.
\citet{2008AJ....136..375M}, who performed detailed abundance studies
of red giants with UVES spectra, also did not confirm the results
of \citet{1999A&AS..135..103A}, who concluded from multicolour CCD
photometry that NGC\,1978 consists of two sub-populations
differing by $0.2$~dex in metallicity. According to their work,
the north-western half of the cluster is more metal-rich and
younger than the south-eastern half (plus differences of the
sub-populations in the helium content). This result has to be seen
in conjunction with a peculiarity of the cluster, namely its high
ellipticity of $\epsilon=0.30$\footnote{The flattening of a
globular cluster is defined as $\epsilon\equiv(a-b)/a$, where $a$
is the major axis and $b$ is the minor axis of the cluster.} which
has been found independently by various groups and was most
recently confirmed by \citet{2007AJ....133.2053M}. A merger event
that could explain both the elongated shape and the suspected two
sub-populations was, however, ruled out for example by
\citet{1992AJ....104.1086F}. A tidal interaction with the host
galaxy seems to be a more likely explanation for the high
ellipticity \citep[][and references therein]{2008AJ....135.1731V}.

\citet{2007AJ....133.2053M} derived the cluster age, distance
modulus, and turn-off mass by fitting isochrones to the cluster
colour-magnitude diagram (CMD). The resulting values are
$\tau=1.9\pm0.1\,\mathrm{Gyr}$, $(m-M)_0=18.50$ and
$M_\mathrm{TO}=1.49\,M_\odot$, respectively, for the best
fit found when using isochrones from the \textit{Pisa Evolutionary
Library} \citep[PEL,][]{2003A&A...404..645C}. When adopting other
isochrones with different overshooting prescriptions, the
parameters vary so that one could also derive a somewhat lower
turn-off mass ($M_\mathrm{TO}=1.44\,M_\odot$), a lower
distance modulus ($(m-M)_0=18.38$), and a cluster age from $1.7$ to
$3.2\,\mathrm{Gyr}$.

The various radial velocity measurements quoted in the literature
widely agree within the error bars, e.\,g.
\citet{1991AJ....101..515O} and \citet{1992AJ....103..447S} derived
$\left<v_\mathrm{r}\right>=+292\pm1.4\,\mathrm{km\,s^{-1}}$
whereas \citet{2006ApJ...645L..33F} determined a mean heliocentric
velocity of NGC\,1978 of
$\left<v_\mathrm{r}\right>=+293.1\pm0.9\,\mathrm{km\,s^{-1}}$ and
with a velocity dispersion $\sigma=3.1\,\mathrm{km\,s^{-1}}$.

NGC\,1978 harbours a number of red giant stars
\citep{1980MNRAS.193...87L} some of which are known to be carbon
stars \citep{1990ApJ...352...96F}. The easiest explanation
for such an occurrence is that these C stars are AGB stars undergoing
the third dredge-up or, alternatively, that they underwent mass accretion from
an old AGB companion, which subsequently evolved into a white dwarf.
More exotic explanations will be discussed in Sect.~\ref{sect:alternativescenarios}.

\subsection{Spectroscopy}

\begin{table*}
\caption{NGC\,1978 targets and log of observations (LE stands for \citealp{1980MNRAS.193...87L})}
\label{table:targets}
\centering
\begin{tabular}{lccccclclc}
\hline\hline
ID & Type & $J$ & $K$               & RA & Dec & H band ($\lambda_\mathrm{c}=15585\,\AA$) & (S/N)$_\mathrm{H}$ & K band ($\lambda_\mathrm{c}=23670\,\AA$) & (S/N)$_\mathrm{K}$\\
   &      & \multicolumn{2}{c}{2MASS} & \multicolumn{2}{c}{J2000} & \\
\hline
 LE4 & M & 12.347 & 11.199 & 05 28 43.72 & $-$66 14 03.7 & 2006 Dec 03: 3 $\times$ 1000 s &  65 & 2006 Dec 05: 3 $\times$ 1000 s &  65 \\
 LE6 & C & 12.090 & 10.707 & 05 28 46.27 & $-$66 13 56.4 & 2006 Dec 07: 3 $\times$ 800 s  &  70 & 2006 Dec 06: 3 $\times$ 800 s  &  55 \\
   A & M & 12.253 & 11.116 & 05 28 46.23 & $-$66 13 24.8 & 2006 Dec 03: 3 $\times$ 1000 s &  55 & 2006 Dec 05: 3 $\times$ 1000 s &  65 \\
 LE3 & C & 11.490 &  9.676 & 05 28 44.49 & $-$66 14 03.9 & 2006 Dec 03: 3 $\times$ 400 s  & 105 & 2006 Dec 05: 3 $\times$ 400 s  & 120 \\
LE10 & M & 12.917 & 11.802 & 05 28 44.44 & $-$66 13 59.9 & 2006 Dec 03: 3 $\times$ 1800 s &  60 & 2006 Dec 05: 3 $\times$ 1800 s &  70 \\
 LE5 & M & 12.527 & 11.387 & 05 28 43.64 & $-$66 13 53.0 & 2006 Dec 03: 3 $\times$ 1200 s &  50 & 2006 Dec 05: 3 $\times$ 1200 s &  60 \\
   B & C & 12.367 & 11.078 & 05 28 43.65 & $-$66 14 09.6 & 2006 Dec 07: 3 $\times$ 700 s  &  50 & 2006 Dec 06: 3 $\times$ 1000 s &  55 \\
 LE1 & C & 12.282 & 10.467 & 05 28 48.50 & $-$66 14 59.9 & 2006 Dec 03: 3 $\times$ 600 s  &  50 & 2006 Dec 05: 3 $\times$ 600 s  & 115 \\
 LE2 & C & 12.967 & 11.266 & 05 28 48.62 & $-$66 15 18.7 & - & - & 2006 Dec 06: 3 $\times$ 1200 s & 65 \\
 LE7 & C & 12.764 & 11.214 & 05 28 47.85 & $-$66 14 44.0 & - & - & 2006 Dec 06: 3 $\times$ 1000 s & 75 \\
 LE8 & M & 13.018 & 11.916 & 05 28 48.47 & $-$66 14 38.7 & 2006 Dec 03: 3 $\times$ 2000 s &  30 & 2006 Dec 05: 3 $\times$ 2000 s &  40 \\
 LE9 & M & 13.315 & 12.259 & 05 28 50.73 & $-$66 14 44.2 & 2008 Mar 19: 4 $\times$ 1800 s &  45 & 2006 Dec 06: 3 $\times$ 2000 s &  50 \\
\hline
\end{tabular}
\end{table*}

\begin{figure}
\centering
\includegraphics[width=0.45\textwidth]{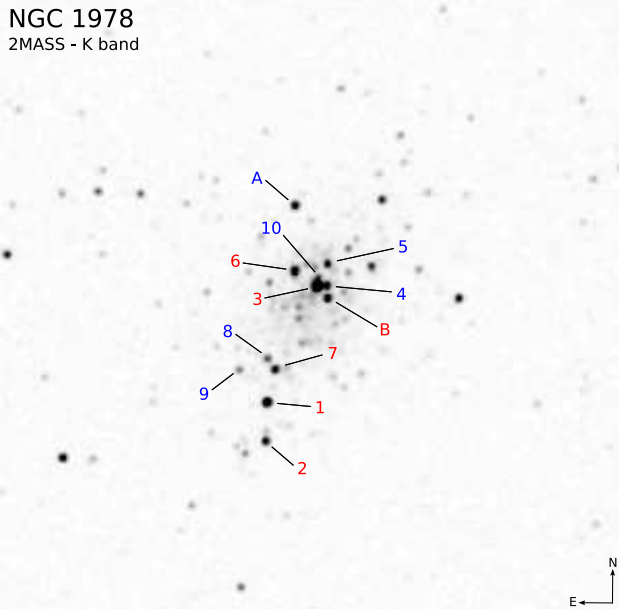}
   \caption{
   Distribution of observed targets in the cluster NGC\,1978. The picture was acquired from the 2MASS catalogue using Aladin \citep{2000A&AS..143...33B}. Numbers refer to the nomenclature of \citet{1980MNRAS.193...87L}. The additional targets A and B were chosen according to their position in the 2MASS colour-magnitude diagram. The stars A, 4, 5, 8, 9, and 10 possess an oxygen-rich atmosphere while B, 1, 2, 3, 6, and 7 are carbon stars. The cluster is elongated and its major axis stretches roughly from south-east to north-west. The displayed sky region is $5\farcm{}0\times5\farcm{}0$ wide. The coordinates of the image centre (close to LE3) read $\alpha=05^\mathrm{h}28^\mathrm{m}45\fs00$, $\delta=-66\degr14'14\farcs{}00$ (J2000).
           }
      \label{fig:ngc1978}
\end{figure}

We recorded high-resolution near-infrared spectra of 12 AGB stars
in the LMC globular cluster NGC\,1978. Ten stars from our target
list were already identified by \citet{1980MNRAS.193...87L} as red
giant stars (and are tagged LE in this work) but largely without
information about the spectral type. \citet{1990ApJ...352...96F}
listed a number of AGB stars in NGC\,1978, comprising also the stars
from LE, and gave information about the spectral type, the cluster
membership, and near-infrared photometry data. Based on this
information, we selected ten targets that could be observed with the
Phoenix spectrograph mounted at the Gemini South Telescope
\citep{1998SPIE.3354..810H,2003SPIE.4834..353H} with reasonable
exposure times. Additionally, we constructed a colour-magnitude
diagram from Two Micron All Sky Survey
\citep[2MASS,][]{2006AJ....131.1163S} data and picked two more stars
that, judging from their $K$ magnitude and $(J-K)$ colour index,
are also located on the AGB of NGC\,1978 (stars A and B).
Details about the observation targets together with the observing log are given in Table~\ref{table:targets}.
The distribution of
all the targets within the cluster is illustrated in
Fig.~\ref{fig:ngc1978}.

Our observing programme required observations at two
different wavelengths, one in the H band and one in the K band. For
this purpose, we utilised the Phoenix order sorting filters H6420
and K4220. However, poor weather conditions in the last of the 4
observing nights prohibited to record H-band spectra for the
targets LE2 and LE7. For the star LE9, we obtained spectra from
queue mode observations in the semester 2008A. The exact
wavelength settings are similar to the ones described in
\citet{2008A&A...486..511L}, i.\,e. in the K band our spectra run
approximately from $23\,580$ to $23\,680\,\mathrm{\AA}$. In the H
band ($15\,530-15\,600\,\mathrm{\AA}$), the spectral region observed was shifted to
slightly higher wavelengths to cover a larger part of the CO 3-0
band head at
$15\,581.6\,\mathrm{\AA}=6417.8\,\mathrm{cm^{-1}}$.
The slit width was set to $0\farcs35$ (the widest slit) which resulted
in a spectral resolution of $R=\lambda/\Delta\lambda=50\,000$.

The total integration time per target and wavelength setting
ranged between 20 (LE3) and 120 (LE9) minutes. For each target we observed 
at two or three different positions along the slit. Each night, we also recorded a
spectrum of a hot star without intrinsic lines in the respective
wavelength regions in order to correct the spectra of our
programme stars for telluric lines. The resulting signal-to-noise 
was from $50$ to above $100$ per resolution element ($\sim4$ pixels).
See Table~\ref{table:targets} for details.

\subsection{Data reduction}
The data reduction procedure was carried out as described at 
length for example in \citet{2002AJ....124.3241S} and in the
Phoenix data reduction IRAF tutorial.\footnote{Available at {\tt
http://www.noao.edu/usgp/phoenix/}. IRAF is distributed by the
National Optical Astronomy Observatories, which are operated by
the Association of Universities for Research in Astronomy, Inc.,
under cooperative agreement with the National Science Foundation.}

To correct the AGB star spectra for telluric lines, we acquired the
spectrum of an early type star without intrinsic lines. The
telluric absorption features were removed using the IRAF task {\tt
telluric} in the K band. The H-band spectra are almost free of
telluric lines, they were, however, also processed in the same way
to remove the fringing. In the H band, we did the wavelength
calibration for a K-type radial velocity standard \citep[HD5457, $v_\mathrm{r}=5.1\,\mathrm{km\,s^{-1}}$,][]{1953QB901.W495.....}
recorded alongside with the programme stars. Using the Arcturus
atlas from \citet{1995iaas.book.....H}, we identified several
features (OH, Fe, Ti, Si lines) in the spectrum and derived a
dispersion solution. The relation was then applied to the
remaining spectra. This procedure allowed us to derive radial
velocities from the H-band spectra. For the K-band spectra we did
a direct calibration using the CO lines in the spectrum of an
M-type target. That solution was then also applied to the carbon-star spectra.

\section{Data analysis}\label{sect:dataanalysis}

\subsection{Contents of the observed wavelength ranges}\label{sect:contents}
The wavelength ranges were chosen such that from the H band
spectra we could derive the stellar parameters and the C/O ratio.
Subsequently, we inferred the carbon isotopic ratio
\element[][12]{C}/\element[][13]{C} from the K band. In practice,
the parameters are tuned iteratively. The region in the H band is
widely used in the literature \citep[e.\,g.][]{2002AJ....124.3241S,2007MNRAS.378.1089M,2008ApJ...689.1020Y}
to derive oxygen abundances from the contained OH lines. From the
relative change of those features in comparison to the band head
from the \element[][12]{C}\element[][16]{O} 3-0 vibration
transition, it is possible to determine the C/O ratio in the
stellar atmosphere. A weak CN line and a few metal lines (Fe, Ti,
Si) aid in the pinning down of the parameters, especially the
effective temperature.

The K-band spectra comprise a number of CO lines from first
overtone ($\Delta\nu=2$) transitions. Beside features from the
main isotopomer \element[][12]{C}\element[][16]{O}, we also find
some \element[][13]{C}\element[][16]{O} lines in this region. We
derived the carbon isotopic ratio by fitting these lines. Apart
from the CO lines, there is also a single HF line blended with a
\element[][13]{C}O feature.

\begin{figure*}
\centering
\includegraphics[width=1.0\textwidth]{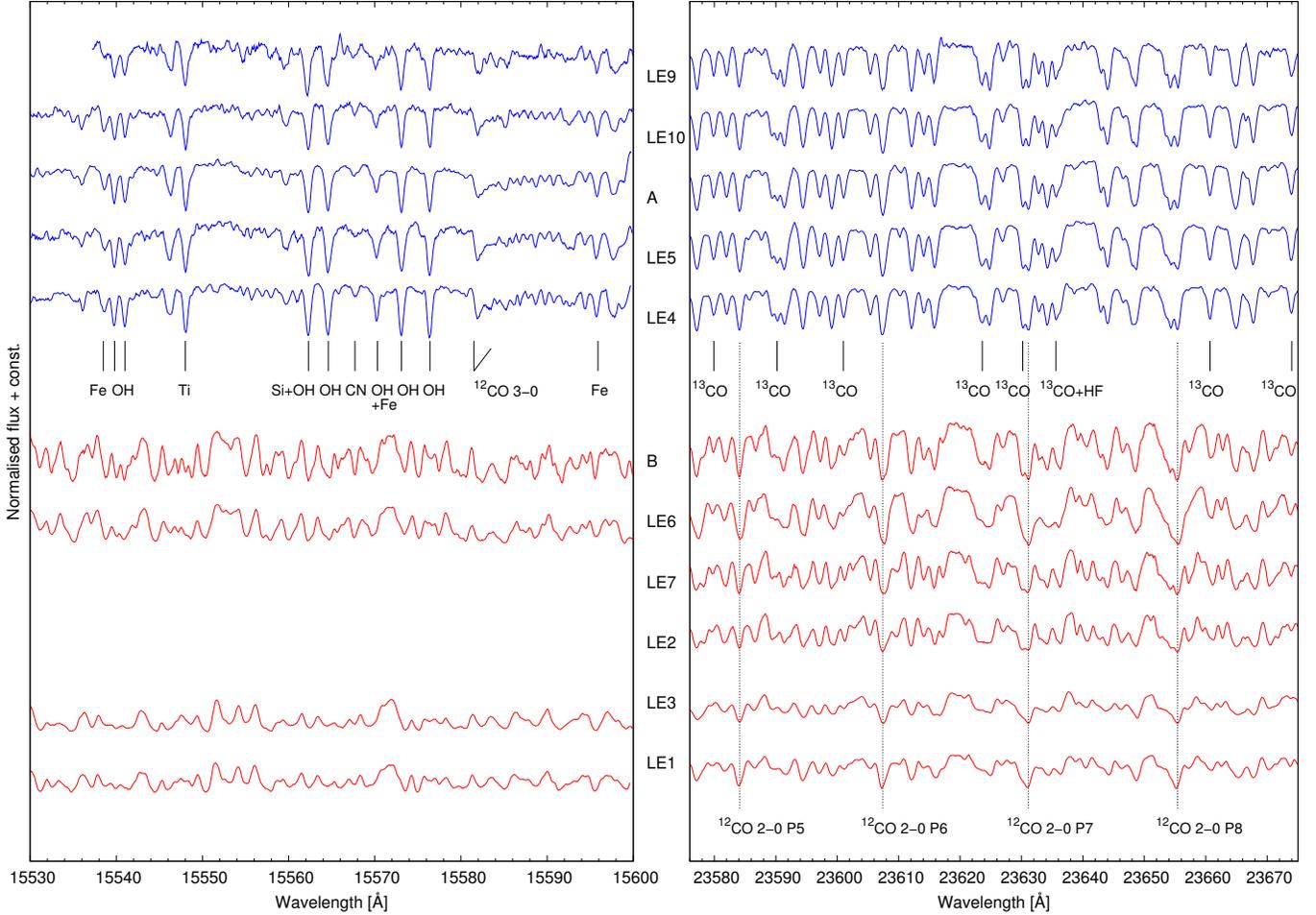}
   \caption{
            Overview about observations: H-band spectra are shown on the left side, K-band spectra on the right. The spectra are ordered by increasing infrared colour $(J-K)_\mathrm{2MASS}$ (from top to bottom), respectively. Some features are identified for the oxygen-rich targets (top), for the carbon-star spectra we only indicate the position of the low-excitation CO lines. Data of the target LE8 are not shown as they are of poor quality and we did not derive abundances from them. H-band spectra for the targets LE2 and LE7 could not be taken due to poor weather conditions. See text for details.
           }
      \label{fig:observations}
\end{figure*}

In Fig.~\ref{fig:observations}, we give an overview about our
observations. For the oxygen-rich stars, the key features in the H
band are marked. In the K band, we indicate the position of the
\element[][13]{C}O lines. The selection of the wavelength ranges was driven by the
oxygen-rich case. From Fig.~\ref{fig:observations} it is obvious
that the regions are not well suited for the analysis of carbon-star 
spectra. Both H- and K-band spectra are crowded with features
from the CN and C$_2$ molecules, occurring in addition to the CO
lines. The polyatomic molecules C$_2$H$_2$, HCN, and also C$_3$
contribute to the opacity by means of many weak lines that form a
pseudo-continuum (see Sect.~\ref{sect:ctypestars}). The major part is due to C$_2$H$_2$ absorption
that increases with decreasing effective temperature. In the H
band, the CO band head is strongly affected by neighbouring
features. Generally speaking, there is not a single unblended line
in the carbon-rich case. The situation in the K band, however, is
not that bad. The \element[][12]{CO} lines are still visible,
although blended with CN features, but there are only a few C$_2$
lines. The pseudo-continuum is essentially all due to C$_2$H$_2$,
contributions from other molecules are negligible.

\subsection{Synthetic spectra}\label{sect:models}
We compare our observations with synthetic spectra based on model
atmospheres that were calculated with the COMARCS code (details
will be given in a forthcoming paper by \citealp{2009arXiv0905.4415A}). 
COMARCS is a modified version of the MARCS code (originating from
\citealp{1975A&A....42..407G,1992A&A...261..263J}, recently
extensively described by \citealp{2008A&A...486..951G}). In the
model calculations, we deduced the temperature and, accordingly,
pressure stratification assuming a spherical configuration in
hydrostatic and local thermodynamic equilibrium (LTE). For stars
showing only a small variability, LTE is a reasonable
approximation. Possible deviations from LTE have a smaller
influence on the abundance determination than large
deviations from the hydrostatic equilibrium (see for example
\citealp{1994LNP...428..234J} for a discussion of non-LTE effects
in cool star atmospheres). In the evaluation of the chemical
equilibrium, which is a consequence of LTE, we take all relevant
opacity sources into account in both model calculation and
spectral synthesis. In the oxygen-rich case H$_2$O, TiO, CO, and
CN are major opacity contributors, while HCN, C$_2$H$_2$, C$_2$ (and others)
are important in the carbon-rich case. The opacity coefficients
utilised by COMARCS are calculated with the COMA code
\citep{Aringer2000,2009arXiv0905.4415A}. The atomic
line data were taken from VALD \citep{2000BaltA...9..590K}, an
overview of the molecular line lists used with all the sources
documented can be found in \citet{LedererAringer2009}. Model
atmospheres and synthetic spectra that were calculated following
the method outlined above have been shown to describe the spectra
of cool giant stars appropriately
\citep[e.\,g.][]{2001A&A...371.1065L,2002A&A...395..915A}. We
applied the same procedure successfully in paper~I for our 
analysis of the AGB stars in the LMC cluster NGC\,1846.

From the parameters that determine an atmospheric model, we held
the mass and the metallicity constant. The respective values are
$M\,=\,1.5\,M_\odot$ and $\mathrm{[M/H]}\,=\,-0.4$, and
were taken from the literature (see Introduction). In the model
calculations, the microturbulent velocity was set to
$\xi\,=\,2.5\,\mathrm{km\,s^{-1}}$, which is in the range that is
found for atmospheres of low-mass giants
\citep[e.\,g.][]{2002A&A...395..915A,2002AJ....124.3241S,2004A&A...422..289G}.
By varying the remaining parameters effective temperature
$T_\mathrm{eff}\,\mathrm{[K]}$, logarithm of the surface gravity
$\log (g\mathrm{[cm\,s^{-2}]})$, and carbon-to-oxygen ratio (C/O),
we constructed a grid of model atmospheres and spectra. This was
done such that we cover the
$T_\mathrm{eff}$ and $\log g$ range resulting from
colour-temperature relations and bolometric corrections
applied to our sample stars. The step
size in effective temperature was set to $50\,\mathrm{K}$, whilst
the surface gravity was altered in steps of $0.25$ on a
logarithmic scale, i.\,e. for $\log g$, ranging from $0.0$ to
$+0.5$. For some carbon stars in our sample we got $\log g<0.0$
from the colour transformations and the adopted mass. The spectral
features used in the analysis show only a minor dependence on this
parameter thus we fixed $\log g=0.0$ in the analysis of the carbon
stars. In this way we also avoid convergence problems for the
atmospheric models.

The element composition is scaled solar
\citep{1993PhST...47..133G}, but we assumed an
oxygen over-abundance of $+0.2\,\mathrm{dex}$ motivated by the
results of our work on NGC\,1846 (paper~I) and by the paper of
\citet{2000A&A...364L..19H}. We altered the C/O ratio in steps of
$0.05$ in the oxygen-rich case and $0.10$ in the carbon-rich case.
This was done by changing the carbon abundance while leaving the
other abundances untouched.

The synthetic spectra cover a wavelength range of
$6400-6450\,\mathrm{cm^{-1}}$ ($15\,504-15\,625\,\mathrm{\AA}$) in
the H band and of $4215-4255\,\mathrm{cm^{-1}}$
($23\,500-23\,725\,\mathrm{\AA}$) in the K band. The spectra were
first calculated with a resolution of
$R=\lambda/\Delta\lambda=300\,000$ and then convolved with a
Gaussian\footnote{The Gaussian has the functional form 
$G(\Delta \lambda,\sigma)=\exp\left\{-(\Delta \lambda/\sigma)^2/2\right\}/(\sigma\sqrt{2\pi})$, 
whereby $\sigma_R\equiv \lambda/(2R)$ to reduce the effective resolution to a given $R$, or
$\sigma_{v_\mathrm{t}}\equiv (\lambda/2)(v_\mathrm{t}/c)$ to account for
the macroturbulent velocity $v_\mathrm{t}$.} 
to match the resolution of our observed spectra
$(R=50\,000)$. By applying another convolution with a Gaussian
profile we took the macroturbulent velocity into account. To
determine the carbon isotopic ratio
\element[][12]{C}/\element[][13]{C} (which is about $89.9$ in the
Sun according to \citealp{1989GeCoA..53..197A}), we altered this
parameter as well. The carbon isotopic ratio has virtually no
effect on the model structure and was consequently only considered
in the spectral synthesis calculations.

\subsection{Determination of abundance ratios}\label{sect:abundances}
We derive the effective temperature $T_\mathrm{eff}$, the surface
gravity $\log g$, the C/O and \element[][12]{C}/\element[][13]{C}
ratio of our targets by fitting synthetic spectra to the
observations. The initial guesses for $T_\mathrm{eff}$ and $\log g$ 
were obtained using colour-temperature relations and
bolometric corrections from the literature. We converted the
$K_\mathrm{2MASS}$ and $(J-K)_\mathrm{2MASS}$ values into the
Johnson system with the formulae given in
\citet{2001AJ....121.2851C}. For the oxygen-rich targets, we
estimated $T_\mathrm{eff}$ and $\log g$ from the relations given
by \citet{2000AJ....119.1424H}. In the case of the carbon stars,
we utilised the work of \citet{1983ApJ...272...99W} and
\citet{1983MNRAS.202...59B} to obtain bolometric corrections
($\mathrm{BC}_K$) and effective temperatures, respectively. From
the $K$ magnitude together with $\mathrm{BC}_K$ and the distance
modulus (see Sect.~\ref{sect:observations}) we derived $\log
L/L_\odot$, and subsequently $\log g$ with the obtained
$T_\mathrm{eff}$ and by assuming $M\,=\,1.5\,M_\odot$.
The reddening of this cluster was taken into account, but it is
low ($E(B-V)=0.05$ and $A_K=0.017$,
\citealp{2007AJ....134..680G}) and has no significant influence on
the transformations.

For each target we fit the parameters in a two-step
process. The idea is to fit the spectral ranges at once rather
than tuning individual abundances (except for carbon) to fit
single spectral features. From the H band, we could in this way
constrain the stellar parameters ($T_\mathrm{eff}$ and $\log g$)
and the C/O ratio. Variations in $\log g$ had the smallest effect
on the synthetic spectra. A small uncertainty in the stellar mass
or radius estimate has thus only a minor influence on the derived
abundance ratios. Using the parameters as obtained from the H
band, we then calculated K-band spectra with a varying
\element[][12]{C}/\element[][13]{C} ratio to determine this
parameter as well. The carbon isotopic ratio does not influence
the appearance of the H band spectrum significantly, as was
verified by test calculations.

To match the shape of the spectral features we also had to assume
a macroturbulent velocity which reduces the effective resolution
of the spectra. This artificial broadening includes the 
instrumental profile, but it is also used to imitate at least partly
the dynamical effects in the stellar atmosphere, which become 
increasingly important for carbon stars. This is also why the
adopted values were generally lower for the
oxygen-rich targets: they were fitted by applying a macroturbulent
velocity of $v_\mathrm{t}=3\,\mathrm{km\,s^{-1}}$. The carbon
stars displayed a stronger broadening of the features (cf.
Fig.~\ref{fig:observations}). We needed values of
$v_\mathrm{t}=8\,\mathrm{km\,s^{-1}}$ and even
$v_\mathrm{t}=10\,\mathrm{km\,s^{-1}}$ (LE6) to fit the spectra
(compare also \citealp{2006A&A...446.1107D} who find comparable 
values in their study). We consistently applied the same value for 
the H and the K band.

We tried to objectify the search for the best fit by applying a
least-square fitting method. Although we were in this way able to
narrow down the possible solutions to a few candidate spectra
quickly, the final decision about our best fit was done by visual
inspection. The reason is that spectral features with a false
strength or position (both due to imperfect line data)
in the synthetic spectrum, or spectral
regions with a high noise level dominate $\chi^2$ and confuse an
automatic minimisation algorithm. As the last step in the fitting
procedure is done by eye, we assign a formal error to the derived
parameters. In the case that the best fit parameters lie
in-between our grid values, we quote the arithmetic mean of the
candidate models as our fit result. The error bars were estimated
from the range of parameters of the model spectra that still gave
an acceptable fit.

The formal errors for the
derived C/O and \element[][12]{C}/\element[][13]{C} ratios 
given in Table~\ref{table:fitresults} compare well with 
the uncertainties found by some basic error estimates, as will be shown
in the following. Therefore the quintessence of the discussion in
Sect.~\ref{sect:discussion} is on a sound footing.
We consider typical uncertainties in the stellar
parameters and investigate the influence on the derived abundance
ratios. The correlations between the stellar parameters and possible systematic errors from the model syntheses have not been taken into account in the error analysis.
For the oxygen-rich case, we start from a baseline model with 
$T_\mathrm{eff}=3750\,\mathrm{K}$, $v_\mathrm{t}=3\,\mathrm{km\,s}^{-1}$, 
$\log g=0.25$, and $\mbox{C/O}=0.15$. Changes of 
$\Delta T_\mathrm{eff}=100\,\mathrm{K}$, $\Delta \log g=0.25$, and 
$\Delta v_\mathrm{t}=2\,\mathrm{km\,s}^{-1}$ 
correspond to changes 
in $\Delta(\mbox{C/O})$ of $0.02$, $0.03$, and $0.01$, respectively\footnote{Note 
that we also include changes of the macroturbulent velocity in the error analysis. 
The equivalent width of a spectral line is unaffected by $v_\mathrm{t}$, but in 
the spectrum synthesis method uncertainties in this parameter influence the
derived abundances.}.
Summed in quadrature this results in a typical total uncertainty of 
$0.04$ for the C/O ratio of our M-type targets. The same exercise
for the carbon isotopic ratio results in $\Delta(\mbox{\element[][12]{C}/\element[][13]{C}})$ of $2$, $2$, and $3$ (parameter dependence as above), and 
thus a total uncertainty of $4$. Concerning the carbon-rich case, we start from a carbon-rich model with parameters $T_\mathrm{eff}=3350\,\mathrm{K}$, $\log g=0.25$, $v_\mathrm{t}=8\,\mbox{km\,s}^{-1}$, $\mbox{C/O}=1.40$, $\mbox{\element[][12]{C}/\element[][13]{C}=175}$ and vary the parameters as described above. This leads to a total $\Delta(\mbox{C/O})=\sqrt{0.1^2+0.05^2+0.1^2}=0.15$. For the carbon isotopic ratio we find $\Delta(\mbox{\element[][12]{C}/\element[][13]{C}})=\sqrt{15^2+15^2+30^2}\simeq37$.

A considerable part of our discussion will be concerned with the
\element[][12]{C}/\element[][13]{C} ratios of our targets.
Since the derived isotopic ratios depend on the
parameters of the \element[][13]{C}O lines \citep[taken from the 
list of][see also Appendix~\ref{sect:colist}]{1994ApJS...91..483G}, we want to assess
possible systematic errors in the line strengths. In the case that
the predicted line strengths in the list are too large, one would
have to increase \element[][12]{C}/\element[][13]{C}
(equivalent to a decrease of the \element[][13]{C} abundance) in
order to fit a \element[][13]{C}O feature compared with the case
where the predicted strengths are correct. The actual
\element[][12]{C}/\element[][13]{C} ratio would then be lower than the derived value. 
For some oxygen-rich stars in our sample we get carbon isotopic ratios that
are close to the value of the CN cycle equilibrium ($4$-$5$). A
necessary further reduction is thus not very likely. If the
theoretical line strengths are too low, the above argument is
reversed. One would assume a lower
\element[][12]{C}/\element[][13]{C} ratio to fit the \element[][13]{C}O
features, which means that one would underestimate the actual isotopic
ratio. Since we, of course, use the same line list for the analysis of
both M and C stars, this would shift up the
\element[][12]{C}/\element[][13]{C} for all targets in
Figs.~\ref{fig:theo1}, \ref{fig:theo2}, and \ref{fig:overshoot}.
Similar to a $\log g$ uncertainty,
the values for the carbon stars
would be stronger affected. Hence, underestimated
\element[][13]{C}O line strengths would relax the necessity of
efficient additional mixing processes on the RGB as discussed
in Sect.~\ref{sect:discussion}.

In the next section, we discuss M and C stars separately. The
features contained in the spectra depend on the chemistry regime
and some issues cannot be discussed in a general way.

\subsubsection{M-type stars}
In the search for a good fit we took advantage of the way features
react to parameter changes. An increasing temperature weakens all
features in the H band. The OH/Fe blend at approximately
$15\,570\,\mathrm{\AA}$ is a good temperature indicator. The two
neighbouring OH lines and the CO band head react less on
temperature changes.
The remaining lines show only a weak
dependence on $T_\mathrm{eff}$. The K band is largely insensitive
to temperature changes, only the \element[][13]{CO} lines show a
weak dependence on temperature which adds to the uncertainty in
determining the isotopic ratio. The only feature strongly reacting
when altering $T_\mathrm{eff}$ in the model calculations
is the blend including the HF line, its strength decreases when
the temperature is increased.

An increase in C/O enhances the strength of the CO band head and
causes stronger CN lines, while the OH lines get slightly weaker,
especially the feature blended with an Fe line. Lowering C/O
decreases the strength of the CO features in the K band.

The CO band head is also sensitive to changes in $\log g$, while
the other features practically are not. 
Changes in the surface gravity also 
affect the measured carbon isotopic ratios (see Sect.~\ref{sect:abundances}).
Two of the \element[][13]{CO} lines in the spectra of the oxygen-rich
stars are unblended (\element[][13]{CO} 2-0 R18 and
\element[][13]{CO} 2-0 R19). Blended \element[][13]{CO} features
were used to check the isotopic ratio derived from the clean
lines.

The continuum is well defined in the oxygen-rich case. We utilised
the least-square method in the process of finding the best fit.

\subsubsection{C-type stars}\label{sect:ctypestars}
Several things make the fitting of carbon stars more difficult
compared with the case of M-type stars. Foremost, the quality of the
line data for the molecules appearing in the carbon-rich case
hampers the qualitative analysis profoundly. For CN, C$_3$, and
C$_2$H$_2$ we used the SCAN database \citep{1997IAUS..178..441J}.
The line positions in the computed CN list are not accurate enough
for modelling high-resolution spectra. With the help of measured
line positions for CN that were compiled by
\citet{2005cns..book.....D}, we were able to correct the
wavelengths of the strongest lines. While the results were
satisfying in the oxygen-rich case, in the carbon-star spectra
many additional weak lines show up that could not be corrected.
The case of C$_2$ \citep[we used the line data from][]{1974A&A....31..265Q}, 
producing a wealth of spectral features, is
even more problematic: no observed reference line list is
available for this molecule, so both line positions and strengths
are subject to large uncertainties. We corrected the line list
manually by removing strong features that did not appear in any of
our observations. Several features were shifted to other
wavelengths where it was evident from the observations that the lines
are at the wrong position. Unlike for M stars, the H-band spectra of carbon
stars are also affected by the carbon isotopic ratio. Lines from
\element[][13]{C}\element[][12]{C} or
\element[][13]{C}\element[][14]{N} are important in some blends,
however, the quality of the line data for these features could not
be assessed for the above described reasons.

The molecules C$_2$H$_2$ and C$_3$ are incorporated into our
calculations via low-resolution opacity sampling data. The
absorption thus becomes manifest as a pseudo-continuum in the
spectra. A possible occurrence of strong lines or regions with
particularly low absorption cannot be reproduced with this
approach. The pseudo-continuum level reacts sensitively to
temperature changes. In general, an increase in $T_\mathrm{eff}$
reduces the feature strength. The relative changes of different
line strength can be used to constrain the temperature range. 
The lower the temperature gets, the higher the contribution of
C$_2$H$_2$ and C$_3$ becomes, whereas C$_2$H$_2$ dominates the
absorption. An increase in the C/O ratio has the same effect.
The implications for abundance determination of the ill-defined
continuum in carbon stars is discussed in detail in paper~I. 
We want to stress here the consequences for the $\chi^2$ method.
The absorption in the pseudo-continuum varies with temperature requiring scaling of
the observed spectrum before a comparison. Due to this scaling the
value of $\chi^2$ changes in the same manner, so that model
spectra with lower effective temperatures (causing a lower flux
level due to an increased pseudo-continuum) always result in a
lower $\chi^2$, pretending to fit the observations better. This is
of course an artificial effect, and thus we cannot rely on the
least-square method as objective criterion.

A change in $\log g$ has only small effects on the carbon-star
spectra. A higher surface gravity produces a higher
pseudo-continuum in the K band.
In general $\log g$ only marginally affects the spectral lines.
However, in the determination of \element[][12]{C}/\element[][13]{C}
the uncertainty in $\log g$ has to be taken into account.
In the K band, all
\element[][13]{CO} lines are blended with other features. The
carbon isotopic ratio is already so high that the strengths of the
\element[][13]{CO} lines have become rather insensitive to changes
in this parameter. As a consequence, even the small uncertainties
in $\log g$ correspond to large changes in the isotopic ratio,
which adds to the errors.

Variations of the C/O ratio have the strongest impact on the H-band 
spectra when C/O is slightly above $1$. The strength of the
CO band head rapidly drops when C/O is increased to $1.3$,
approximately. Then a saturation sets in and the strength of the
CO band head varies slowly with C/O. This behaviour and strong CN
features sitting in the band head hamper an accurate C/O
determination. The lines of C$_2$ and CN increase in strength when
C/O rises, both in the H and the K band. The CO lines in the K
band decrease in strength for higher values of C/O. The changes
are, however, small and do not allow for a determination of the
C/O ratio from the K-band spectra alone.

\begin{figure*}[t]
\centering
\includegraphics[width=\textwidth]{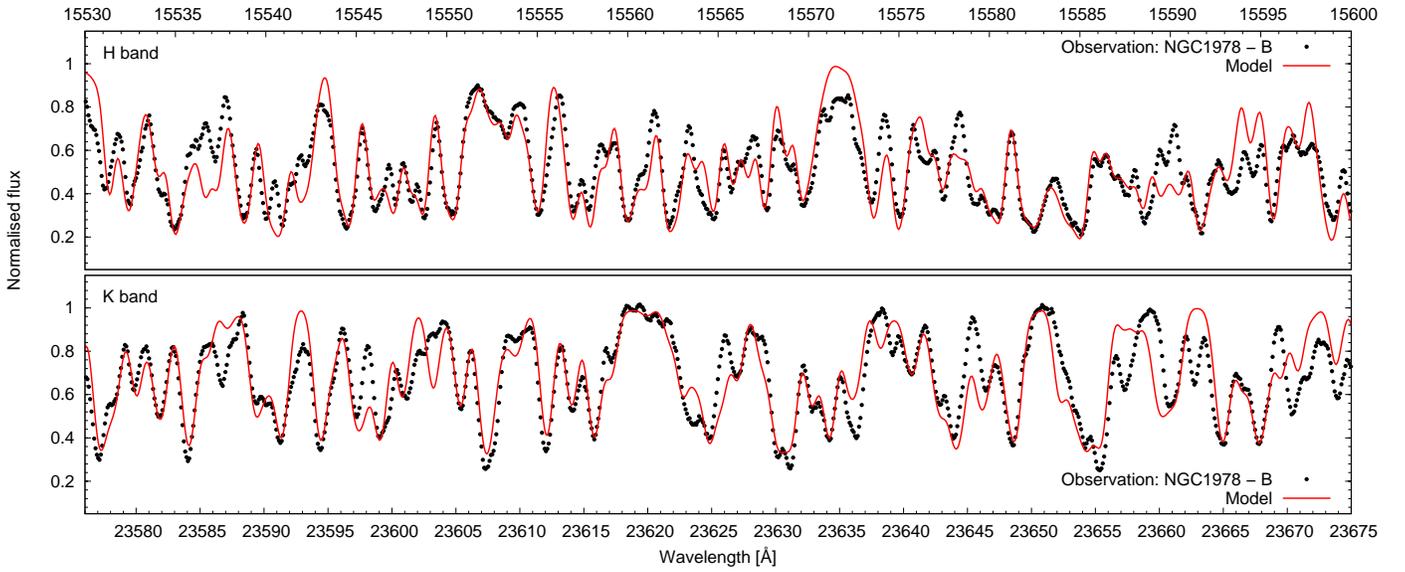}
   \caption{Observation and fit of the carbon star B in NGC\,1978. The fit parameters are given in Table~\ref{table:fitresults}. The relatively high effective temperature and low C/O ratio compared with the other carbon stars in our sample allowed a reasonable fit. Deviations can to a high fraction be ascribed to uncertain line data. The strong low-excitation CO lines in the K band are not reproduced by the models. See text for details.
           }
      \label{fig:fitexample}
\end{figure*}

Figure~\ref{fig:fitexample} is, as an example, a fit for the
star B, which is a carbon star. The star has an effective
temperature of about $3350\,\mathrm{K}$ (see also
Table~\ref{table:fitresults} for the other fit parameters). The
pseudo-continuum contribution is relatively weak, although there
are almost no line-free regions in both spectral bands. A number
of features is successfully reproduced by our models, the
deviations in other regions are most probably due to uncertain
line data. 

To measure the \element[][12]{C}/\element[][13]{C} ratio in the carbon-star 
spectra, we utilised the \element[][13]{C}O lines at $23\,579.9$, 
$23\,590.2$, and $23\,601.0\,{\rm \AA}$ (the leftmost lines indicated in the 
right panel of Fig.~\ref{fig:observations}). In the vicinity of the other 
\element[][13]{C}O lines, there are obviously absorption features missing 
in our synthetic spectra (see Fig.~\ref{fig:fitexample}). We cannot rule 
out that our analysis of a limited number of line blends could introduce 
a systematic error in the inferred carbon isotopic ratios. Therefore, we 
emphasise that it would be worthwhile to reassess the isotopic ratios in other 
wavelength regions, particulary in the light of the discussion in 
Sect.~\ref{sect:discussion}.

We also identified four low-excitation \element[][12]{C}O 2-0 lines 
in the K band (see also Fig.~\ref{fig:observations}). 
According to \citet{1982ApJ...252..697H}, low-excitation lines form 
in the outer atmospheric layers, opposite to high-excitation lines 
that show characteristics similar to second overtone transitions
($\Delta\nu=3$). In all the carbon-star spectra, these lines could
not be fitted with our synthetic spectra. The calculations
underestimate the line strength, suggesting that the extended
atmospheres of luminous carbon stars are not well reproduced by
our hydrostatic models.

\subsection{Stellar evolutionary models}\label{sect:evolutionarymodels}
The stellar evolutionary models presented in this paper have
been computed with the FRANEC code \citep{1998ApJ...502..737C}.
The release we are currently using is optimised to properly
compute low-mass models during their AGB phase. Up-to-date input
physics, such as low-temperature carbon-enhanced opacities, have been
adopted \citep{2007ApJ...667..489C,LedererAringer2009}. 
Physical phenomena, such as hydrodynamical instabilities at
radiative--convective interfaces and the mass-loss rate, have been
properly taken into account \citep{2006NuPhA.777..311S,CristalloStranieroGallinoPiersantiDominguezLederer2009}. The inclusion of an additional mixing mechanism taking place below the convective envelope (usually referred to as {\em extra-mixing})  during the RGB and the AGB phase was described in paper~I.

\section{Results}\label{sect:results}

\subsection{Cluster membership}
We derived the heliocentric radial velocity by cross-correlating
the H-band spectra with a template spectrum (using the IRAF task {\tt
fxcor}). Data for the star LE9 were taken in the semester 2008A,
and no radial velocity standard (K-type star) was recorded, so no
information could be deduced for this target. While the quality of
our LE8 data is too low to derive abundance ratios, it is still
adequate to measure the radial velocity (only with a
slightly larger error than for the other targets). The sample
shows a relatively homogeneous radial velocity distribution, the spread is
quite narrow, so we conclude that all our targets are actually
cluster members. The results are summarised in
Table~\ref{table:radialvelocities}. The value
$v_\mathrm{r}=+292.9\pm1.3\,\mathrm{km\,s^{-1}}$ (with a velocity
dispersion of $\sigma_v=2.6\,\mathrm{km\,s^{-1}}$) we find is well
in line with earlier determinations (see Introduction). As a comparison,
\cite{2008AJ....135..836C} measured radial velocities in four fields with
different distances to the LMC centre. The mean values found
in the individual fields range from $+278$ to $+293\,\mathrm{km\,s^{-1}}$.
The velocity dispersion in the metallicity bin with ${\rm [Fe/H]\geq-0.5}$ is 
$\sigma_v=20.5\,\mathrm{km\,s^{-1}}$.

The errors for the carbon stars are systematically larger than for the
oxygen-rich stars. This can easily be understood by looking at
Fig.~\ref{fig:observations}. The features in the C-rich case are
usually broader and consequently the peak in the cross-correlation
function (and thus the FWHM) is broader, too.

\begin{table}
\caption{Radial velocities of NGC\,1978 targets}
\label{table:radialvelocities}
\centering
\begin{tabular}{lccc}
\hline\hline
ID & $v_\mathrm{r} \mathrm{[km\,s^{-1}]}$ & & \\
\hline
A    & $+295.6$ & $\pm$ & $1.0$\\
LE4  & $+296.4$ & $\pm$ & $0.8$\\
LE5  & $+292.9$ & $\pm$ & $0.9$\\
LE8  & $+290.4$ & $\pm$ & $1.3$\\
LE10 & $+292.5$ & $\pm$ & $0.7$\\
\\
B    & $+290.6$ & $\pm$ & $1.3$\\
LE1  & $+297.5$ & $\pm$ & $1.6$\\
LE3  & $+291.5$ & $\pm$ & $1.9$\\
LE6  & $+290.2$ & $\pm$ & $1.7$\\
\hline
mean                & $\mathbf{+292.9}$ & $\mathbf{\pm}$ & $\mathbf{1.3}$ \\
velocity dispersion ($\sigma$) &          &     & $2.6$ \\
\hline
\end{tabular}
\end{table}

\subsection{Stellar parameters and abundance ratios}

\begin{table*}
\caption{Data and fit results for oxygen-rich (first group) and carbon-rich (second group) targets in NGC\,1978}
\label{table:fitresults}
\centering
\begin{tabular}{lrrcr|ccrcc}
\hline \hline
ID & $K$ & $J-K$ & $T_\mathrm{eff,c}$ & $L/L_\odot$ & $T_\mathrm{eff,f}$ & $\log g$ & $v_\mathrm{t}$ & C/O & $^{12}\mathrm{C}/^{13}\mathrm{C}$ \\
& \multicolumn{2}{c}{2MASS} & [K] & & [K] & & $\mathrm{[km\,s^{-1}]}$ & & \\
\hline
A    & 11.116 & 1.137 &  3600 & 5300  & 3825 & 0.50 &  3 & $0.23\pm0.05$ & $16\pm3$ \\
LE4  & 11.199 & 1.148 &  3600 & 5000  & 3725 & 0.38 &  3 & $0.18\pm0.03$ & $13\pm4$ \\
LE5  & 11.387 & 1.140 &  3600 & 4200  & 3775 & 0.25 &  3 & $0.18\pm0.05$ &  $6\pm2$ \\
LE9  & 12.259 & 1.056 &  3800 & 2000  & 3900 & 0.38 &  3 & $0.13\pm0.03$ & $12\pm2$ \\
LE10 & 11.802 & 1.115 &  3700 & 3300  & 3900 & 0.38 &  3 & $0.18\pm0.05$ &  $9\pm3$ \\
& & & & & & & & & \\
B    & 11.078 & 1.289 &  3176 & 4400  & 3350 & 0.00 &  8 & $1.35\pm0.10$ & $175\pm25$ \\
LE1  & 10.467 & 1.815 &  2554 & 6200  & 2600 & 0.00 &  8 & $>1.50$       & $>200$     \\
LE3  &  9.676 & 1.814 &  2556 & 12900 & 2600 & 0.00 &  8 & $>1.50$       & $>200$     \\
LE6  & 10.707 & 1.383 &  3043 & 5800  & 3100 & 0.00 & 10 & $1.30\pm0.10$ & $150\pm50$ \\
\hline
\end{tabular}
\end{table*}

We summarise our fit results in Table~\ref{table:fitresults}. In the
first column, we list the star identifier (compare
Fig.~\ref{fig:ngc1978}), followed by the $K$ magnitude and the
colour index $(J-K)$ taken from the 2MASS database. In the next
two columns, we quote the effective temperature and luminosity
(rounded to $100\,L_\odot$) resulting from the respective
colour calibrations and bolometric corrections (see
Sect.~\ref{sect:abundances}). The parameters resulting from our
fitting procedure ($T_\mathrm{eff}$, $\log g$, $v_\mathrm{t}$,
C/O, \element[][12]{C}/\element[][13]{C}) are listed in the
subsequent columns.
We note that none of our target stars is
a large amplitude variable, thus the influence of variability on
the stellar parameters can be safely ignored. A detailed
discussion of the small amplitude variability ($1.5\,\mathrm{mag}$
in $R$) will be given elsewhere (Wood \& Lebzelter, in
preparation).

\begin{figure*}
\centering
\includegraphics[width=\textwidth]{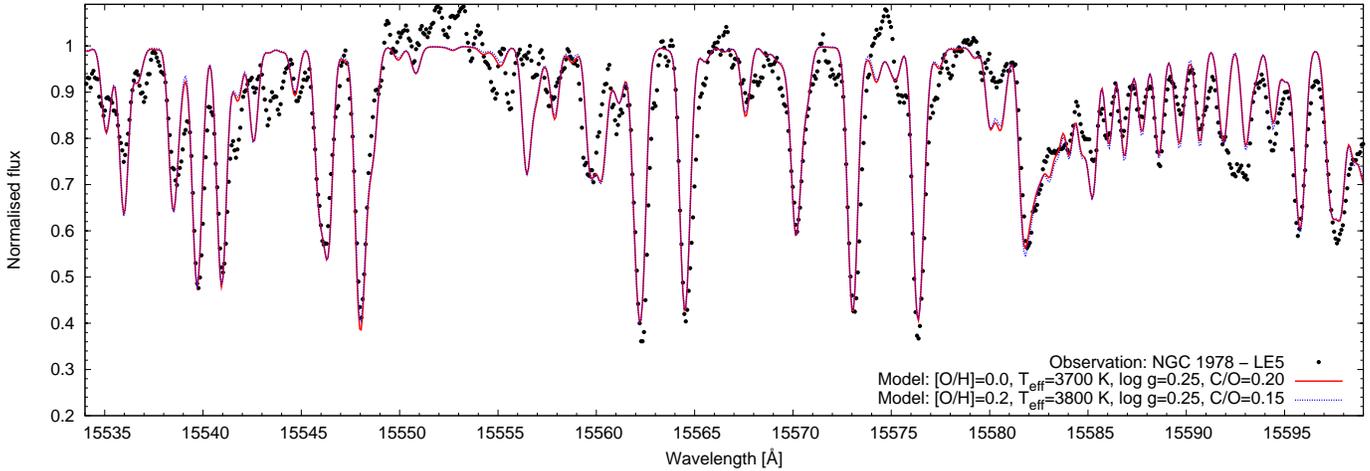}
   \caption{Fit of the H band spectrum of the target LE5 adopting different oxygen abundances. A value of $\mathrm{[O/Fe]}=0.2$ (which we used in our analysis) leads to a higher temperature and a lower C/O ratio compared with a scaled solar oxygen abundance ($\mathrm{[O/Fe]}=0.0$). The two model spectra both give reasonable fits to the observations, thus a decision about the actual oxygen abundance cannot be made from our data. However, lower effective temperatures (and thus the ``no oxygen over-abundance" scenario) agree better with the values derived from colour-temperature relations.
           }
      \label{fig:fitambiguity}
\end{figure*}

The derived effective temperatures are systematically higher than
the $T_\mathrm{eff}$ values inferred from the infrared colour
transformations (compare the resulting $T_\mathrm{eff,c}$ to 
$T_\mathrm{eff,f}$ deduced from the fit in Table~\ref{table:fitresults}).
Better agreement between the two temperature
scales is found when we adopt a scaled solar oxygen abundance
rather than an over-abundance of $0.2\,\mathrm{dex}$, which was
our standard choice in the analysis (we refer to
Sects.~\ref{sect:introduction} and \ref{sect:models} for details).
However, this is not too surprising since the colour-temperature
relations from \citet{2000AJ....119.1424H} were derived from synthetic
spectra based on scaled solar abundances.
We want to mention that the results from
\citet[][their Fig.~8]{2002AJ....124.3241S} imply
$\mathrm{[O/Fe]}\le0.0$ for the LMC, contrary to our assumption.
Also, \citet{2008AJ....136..375M} found that the other alpha element
abundances are roughly scaled solar
($\mathrm{[\alpha/Fe]}\simeq0.0$) in NGC\,1978. 
However, in another recent work \citet{2009AJ....137.4988G}
derived $\mathrm{[\alpha/Fe]}=0.2$ for the LMC cluster NGC\,1846.
Apart from the
influence on the temperature scale, a higher oxygen abundance
would shift the derived C/O ratios only to slightly higher
values (Fig.~\ref{fig:fitambiguity}). The derived carbon isotopic
ratios are marginally affected by an oxygen over-abundance.

\begin{figure}
\centering
\includegraphics[width=0.48\textwidth]{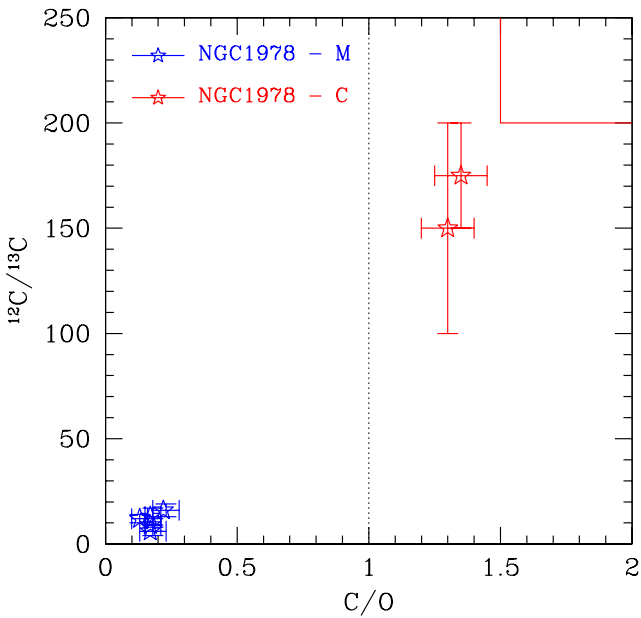}
   \caption{
            Measured abundances ratios C/O and \element[][12]{C}/\element[][13]{C} for the targets in NGC\,1978. Oxygen-rich targets appear to the left of the dotted line at $\mbox{C/O}=1$, carbon-rich targets are on the right. The box in the upper right corner marks the lower limits for the targets LE1 and LE3. There are no obvious signs of TDU among the M stars (see left panel of Fig.~\ref{fig:theo1} for a zoom to the M-star data points), S stars are completely lacking in our sample of cluster stars. Additionally, from the carbon-star data we cannot detect a saturation level for the carbon isotopic ratio which would hint at the occurrence of extra-mixing.
           }
      \label{fig:fitresults}
\end{figure}

We found that for the five oxygen-rich stars within our sample the
C/O ratio is varying little, the values range from $0.13$ to
$0.18$ with a typical uncertainty of $\pm0.05$. The isotopic
ratios vary in the range between $9$ and $16$ with an uncertainty
ranging up to $\pm4$. Considering the error bars, all M stars
occupy more or less the same region in Fig.~\ref{fig:fitresults}, where
we display the measured C/O and \element[][12]{C}/\element[][13]{C} of
our sample stars.
No target is offset from the spot marking the abundance ratios
expected after the evolution on the first giant branch
($\mbox{C/O}\simeq0.2$,
\element[][12]{C}/\element[][13]{C}$\simeq10$). In other words, we
do not find conclusive signs of third dredge-up among the
oxygen-rich stars in the cluster. This is also consistent with the
results from \citet{1983MNRAS.204..985E} who did not find S-type
stars in NGC\,1978. The total number of our targets is small, so general statements
based on our results are rather weak due to the poor statistics. 
However, the fact that we find oxygen-rich stars, in which the TDU
does not seem to be active, does not exclude that this phenomenon is
at work in other stars of the cluster. 

In fact, we also identify a sub-sample of
carbon-rich stars belonging to NGC\,1978. Satisfactory fits could only
be achieved for the two hottest carbon stars (B and LE6) in our
sample. We also derived C/O and
\element[][12]{C}/\element[][13]{C}. 
The error bars are larger than for the M stars, the reasons 
for that are outlined in the previous sections.
For the two cool objects LE1 and LE3, we could
only derive lower limits for C/O and
\element[][12]{C}/\element[][13]{C}. For an
increasing carbon content, the features saturate, thus we cannot give a
reliable upper limit for C/O. 
The C/O ratios that we found are in line with the
results obtained by other groups. \citet{2005A&A...434..691M} adopted $\mbox{C/O}\geq1.4$ 
to explain molecular features in low-resolution spectra of LMC carbon stars.
Investigations of planetary nebulae in the LMC exhibit a range 
of C/O ratios from
slightly above $1$ up to even $5$ \citep[e.\,g.][]{2005ApJ...622..294S}.
We are not aware of measurements of the carbon isotopic ratio
directly from carbon-rich AGB stars in the LMC
other than the ones presented here and in paper~I. 
\citet{2007A&A...461..641R} analysed the post-AGB object 
MACHO~47.2496.8 ($\mathrm{[Fe/H]}=-1.4$) in the LMC and derived $\mbox{C/O}>2$ and 
$\element[][12]{C}/\element[][13]{C}\sim200$. A peculiar combination of 
C/O and \element[][12]{C}/\element[][13]{C} values even more extreme than those of 
our carbon-rich targets was presented by \citet{2006A&A...446.1107D}. 
They found $\mbox{C/O}\leq1.2$ and $\element[][12]{C}/\element[][13]{C}>300$ for 
{BMB-B~30} (with $\mathrm{[M/H]}=-1.0$) in the SMC.

Qualitative estimates based on the K-band spectra of the targets 
LE2 and LE7 suggest that those have an
effective temperature between the groups LE1--LE3 and B--LE6 
(see Fig.~\ref{fig:observations}). Except for the target
B, the difference between the derived effective temperature and 
the $T_{\rm eff}$ value obtained from the colour transformation is not quite as
high as in the oxygen-rich case (about $50\,\mathrm{K}$). We
repeated the analysis for the star B with a scaled solar oxygen
abundance (as for LE5, Fig.~\ref{fig:fitambiguity}) and also found
a lower $T_\mathrm{eff}$, whereas the C/O ratio remained basically
unchanged. Almost all spectral features in the carbon-rich case react
sensitively to temperature changes, therefore $T_\mathrm{eff}$ can
be better defined than the C/O ratio. A remarkable result is that
---opposite to our findings for NGC\,1846 (paper~I)---we do not
find a saturation level for the carbon isotopic ratio. This
indicates that {\em extra-mixing is not at work in NGC\,1978}.

\begin{figure}
\centering
\includegraphics[width=0.48\textwidth]{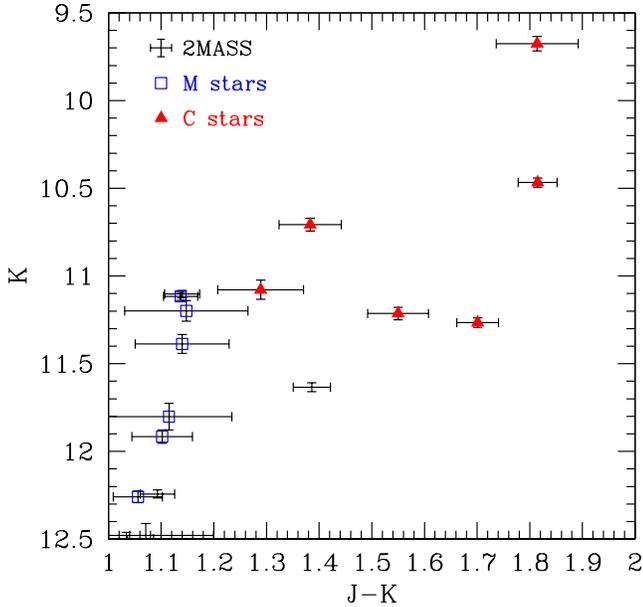}
   \caption{Colour-magnitude diagram based on 2MASS data for NGC\,1978 (complete within a radius of $4\farcm{}0$ with respect to the cluster centre). Empty squares refer to oxygen-rich objects in our sample, filled triangles depict carbon stars. Compare also Table~\ref{table:fitresults}. Details are discussed in the text.
           }
      \label{fig:cmd}
\end{figure}

In Fig.~\ref{fig:cmd} we show a CMD of NGC\,1978 (see 
Sect.~\ref{sect:discussion} for a discussion of the photometric errors).
We do not plot the bolometric magnitudes on the
ordinate because of the inconsistent bolometric corrections
applied to the targets.
The CMD contains all targets that lie within a radius of
$4\farcm{}0$ with respect to the cluster centre (cf. Fig.~\ref{fig:ngc1978}).
Data points with superimposed symbols refer to stars contained in our sample.
The oxygen-rich stars (empty squares) seem to form a
sequence with a C/O ratio increasing with luminosity which might
hint at a mild dredge-up. However, this could be misleading, as we have
to take the error bars quoted in Table~\ref{table:fitresults} into
account. Moreover, the isotopic ratios do not appear in an ordered
sequence. The carbon stars (red triangles) are brighter in $K$ and have
significantly redder colours. For the stars LE1 and LE3, having
the highest $(J-K)$ values, the indicated numbers refer to lower
limits for the abundance ratios. The two carbon stars 
with the lowest $K$ magnitude are
LE2 and LE7 which were excluded from our analysis. Both have
luminosities comparable to the brightest oxygen-rich
stars. A possible explanation is that these stars are in a
post-flash dip phase \citep{1983ARA&A..21..271I}, displacing them
about $1$ magnitude down from the average luminosity in this
evolutionary stage in the CMD (see also the discussion in
paper~I).

\section{Discussion}\label{sect:discussion}
We seek to explain the following observed features of NGC\,1978:
\begin{enumerate}
\item The cluster harbours carbon stars. \item The M stars do not
show conclusive signs of third dredge-up. \item There is no
saturation of the carbon isotopic ratio.
\end{enumerate}

Except for the first point, this is more or less the opposite of
what we found for the cluster NGC\,1846 (paper~I), the composition of which
is similar to the one of NGC\,1978,
even if the GC is slightly younger
and, as a consequence, harbours more massive AGB stars
(with a mass of about $1.8\,M_\odot$). The M
stars in NGC\,1846 span a range in C/O from $0.2$ to $0.65$, and
\element[][12]{C}/\element[][13]{C}---initially rising together with
$\mbox{C/O}$---does not exceed $60$ to $70$, even in the carbon
stars.

The first point in the list is a clear indication of
ongoing TDU in the cluster. The second point may at first seem to
contrast this statement, while low number statistics could resolve
this apparent contradiction: the lifetime of a thermally pulsing
AGB star is short, so in a sample of a few stars we do not
necessarily have to find a star with an enhanced carbon abundance.
This possibly explains the large gap between the M and the C stars
in terms of C/O (see Fig.~\ref{fig:fitresults}). An alternative
scenario where the TDU is so efficient that the star becomes
carbon-rich after a single dredge-up episode occurs only at much
lower metallicity \citep[e.\,g.][]{2000MmSAI..71..745H}.

Let us put aside the C/O error bars for a moment and speculate
about the group of M stars. The 2MASS data come with uncertainties
in $K$ of lower than $0.05$, except for LE10 where the error is
$0.076$.\footnote{The errors for $(J-K)$ are higher, ranging up to
and being even greater than $0.1$. This could also affect the
differences between the effective temperature scales discussed in
Sect.~\ref{sect:results}.} In any case, the luminosity sequence
that the stars form is conserved. The star LE9 is the faintest at $K$, with $m_K=12.3$,
in the sample and also the star with the lowest C/O ratio.
\citet{2000A&A...359..601C} identify the RGB tip in the LMC at
$K_0=12.1\,\mathrm{mag}$. According to this value, LE9 could as
well be an RGB star or located at the early-AGB. The targets LE5
and LE10 are possibly at the very beginning of the thermally
pulsing asymptotic giant branch (TP-AGB) star phase. The case of
LE4 and, in particular, A is more difficult. Comparing these stars
with those of similar luminosity in the sample of NGC\,1846, we
note that the latter are classified as S stars and present C/O
ratios of $0.4$ or even higher. The C/O ratio we find for the star
A ($0.23$) is slightly higher than the average of the other stars,
perhaps compatible with just one (the first) TDU episode. 
The bolometric magnitude of A also roughly corresponds to the 
limit we found for the onset of the TDU in NGC\,1846 ($m_\mathrm{bol}=14.1\,\mbox{mag}$). 

\subsection{Stellar evolutionary models}
\begin{figure*}
\includegraphics[width=0.48\textwidth]{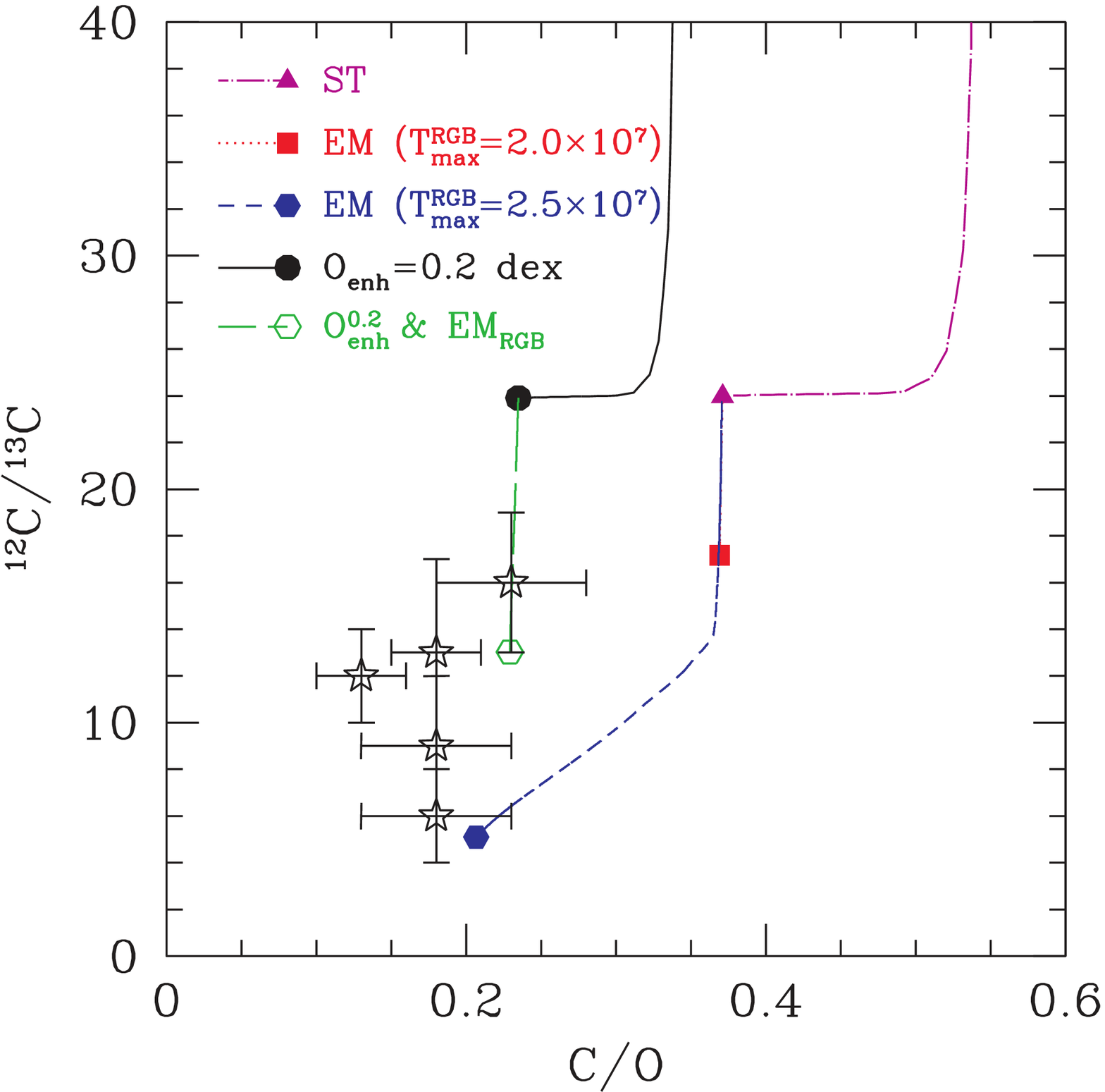}
\includegraphics[width=0.48\textwidth]{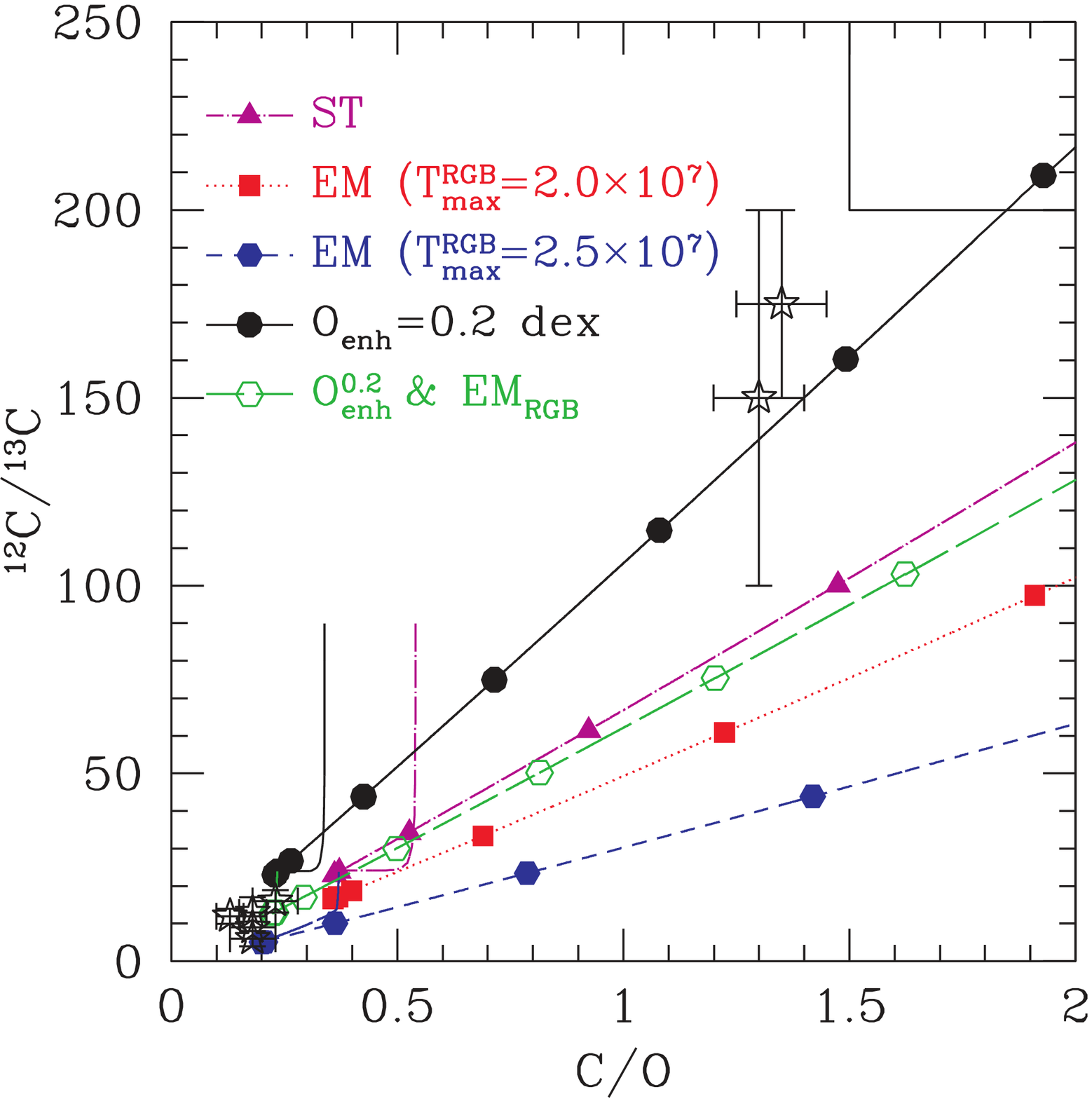}
\caption{Comparison between observational data and theoretical
models. \element[][12]{C}/\element[][13]{C} isotopic ratios vs.
C/O ratios are reported. In the left panel we report the range of
O-rich stars only, while in the right panel the whole sample has
been considered. The star symbols with error bars indicate our observations. 
The red dotted, the blue dashed, and the green long-dashed
curves in the left panel are partly covered by other tracks in the plot. The symbols in the left panel mark the end point of the tracks after the RGB phase. In the right panel, the symbols indicate the values attained during the interpulse phases, which are on the order of $10^3$ times longer than the TDU episodes. See text for details.}
      \label{fig:theo1}
\end{figure*}

In this section we present theoretical tracks computed
with the FRANEC code, and we compare them with observations. In the
left panel of Fig.~\ref{fig:theo1}, we 
report the evolution of the \element[][12]{C}/\element[][13]{C} versus C/O
curves from the pre-main sequence up to the early-AGB phase in the region 
where the M stars of NGC\,1978 lie. In practice, the end point
of each curve (marked by a symbol in the left panel of Fig.~\ref{fig:theo1}) represents
the value attained at the tip of the RGB phase, which is
conserved up to the onset of the first thermal pulse. In the right
panel, instead, we extend the axes to also include the C-rich
stars. We plot the whole AGB evolutionary tracks and mark
the values attained after each TDU episode.

We firstly concentrate on the O-rich stars of our sample. We carry
on our analysis under the assumption that these stars have not
experienced TDU, because they still are at the beginning of the
TP-AGB phase or because they arrived on the AGB with a too small
envelope mass. In this case, their initial surface composition
has been modified by the occurrence of the First Dredge-up
(FDU) only and, eventually, by an additional slow mixing operating
below the convective envelope during the RGB phase (the so called
{\em extra-mixing}, see e.\,g. \citealp{2003ApJ...582.1036N,2007A&A...467L..15C,2008ApJ...684..626D}). Our reference model (ST, dash-dotted magenta curve) has an initial mass $M=1.5\,M_\odot$
with $Z=0.006$, corresponding to $\mathrm{[Fe/H]}=-0.36$. We assume an
initial solar-scaled composition, which implies $\mbox{C/O}=0.5$ and
$\element[][12]{C}/\element[][13]{C}=90$. After the FDU, the surface 
values attain $\mbox{C/O}=0.36$
and $\element[][12]{C}/\element[][13]{C}=24$. This is due to the fact that the
convective envelope penetrates into regions where partial
hydrogen-burning occurred before. No extra-mixing has been included
in the ST case.
The values so obtained clearly disagree with the M stars observations, 
which show an average $\mbox{C/O}=0.18$
and an average $\element[][12]{C}/\element[][13]{C}=11$. Thus, we explored the
possibility of an occurrence of extra-mixing on the RGB. 

This hypothesis is supported by the bulk of observations of RGB stars in the
galactic field, as well as in open and globular 
clusters \citep[see e.\,g.][]{2000A&A...358..671G}.
These observations show that this additional mixing occurs in low-mass stars ($M<2\,M_\odot$)  
during the first ascent along the Red Giant Branch. 
Moreover, \citet{2008ApJ...677..581E} identified a mixing mechanism driven 
by a molecular weight inversion ({\em $\delta\mu$-mixing}) in three-dimensional 
stellar models that must operate in all low-mass stars while they are on the RGB.
The operation of this RGB 
extra-mixing is also required to explain the relatively low 
\element[][12]{C}/\element[][13]{C} ratios in the M stars of NGC\,1846 (paper~I).

As in paper~I, we include this additional mixing right
after the RGB luminosity bump. The extension of the zone in which
this additional mixing takes place is fixed by prescribing the
maximum temperature the material is exposed to
($T^\mathrm{max}_\mathrm{RGB}$). The circulation rate is tuned by setting the
mixing velocity to a value that is a small fraction (cf. paper~I)
of the typical convective velocities in the envelope of an RGB star. As discussed in
paper~I, the carbon isotopic ratio largely depends on
$T^\mathrm{max}_\mathrm{RGB}$. In Fig.~\ref{fig:theo1}
we report two models, characterised by
$T^\mathrm{max}_\mathrm{RGB}=2.0\times10^7$\,K (red dotted curve) and
$T^\mathrm{max}_\mathrm{RGB}=2.5\times10^7$\,K (blue short-dashed curve). In the
first case, the final \element[][12]{C}/\element[][13]{C} ratio decreases, reaching a
value in good agreement with those observed in the M stars of NGC\,1978. Notwithstanding, the C/O ratio remains unaltered and higher
than the observed one. An increase of $T^\mathrm{max}_\mathrm{RGB}$ up to
$2.5\times10^7$\,K leads to a lower C/O ratio of about $0.21$, which is in
better agreement with the observations. However, the corresponding
\element[][12]{C}/\element[][13]{C} ratio is $5$, which is definitely 
lower than the average value ($11$). Note that only LE5 shows such a 
low value ($\element[][12]{C}/\element[][13]{C}=6$).

Then, we explored the possible effect of an oxygen
enhancement (see the discussion in Sect.~\ref{sect:targetcluster}). The
black solid line in Fig.~\ref{fig:theo1} represents a model similar to
the ST case, but with $\mathrm{[O/Fe]}=0.2$. As for the ST case, the effects
of the FDU are clearly recognisable. An additional model, as
obtained by including an RGB extra-mixing
($T^\mathrm{max}_\mathrm{RGB}=2.1\times10^7$\,K) is also reported (green line).
The final surface composition of this model ($\mbox{C/O}=0.23$ and
$\element[][12]{C}/\element[][13]{C}=13$) is close to the average values of the
observed sample ($\mbox{C/O}=0.18$ and $\element[][12]{C}/\element[][13]{C}=11$). Thus, a first
conclusion is that moderate RGB extra-mixing
($2.1\times10^7<T^\mathrm{max}_\mathrm{RGB}<2.3\times10^7$\,K) coupled to
moderate oxygen enhancement appears to nicely reproduce the
observed average composition of the M stars in NGC\,1978. The same
conclusion has been reached in the case of NGC\,1846 (paper~I).

What are the implications of this scenario when applied to the
whole observational sample (O-rich and C-rich stars)?
We carry on our analysis assuming that the carbon stars are intrinsic,
i.\,e. that their surface composition, in particular the high C/O and 
carbon isotopic ratios, is the result of nucleosynthesis
and mixing processes occurring during the thermally pulsing AGB
evolution.

\begin{figure*}[t]
\includegraphics[width=0.48\textwidth]{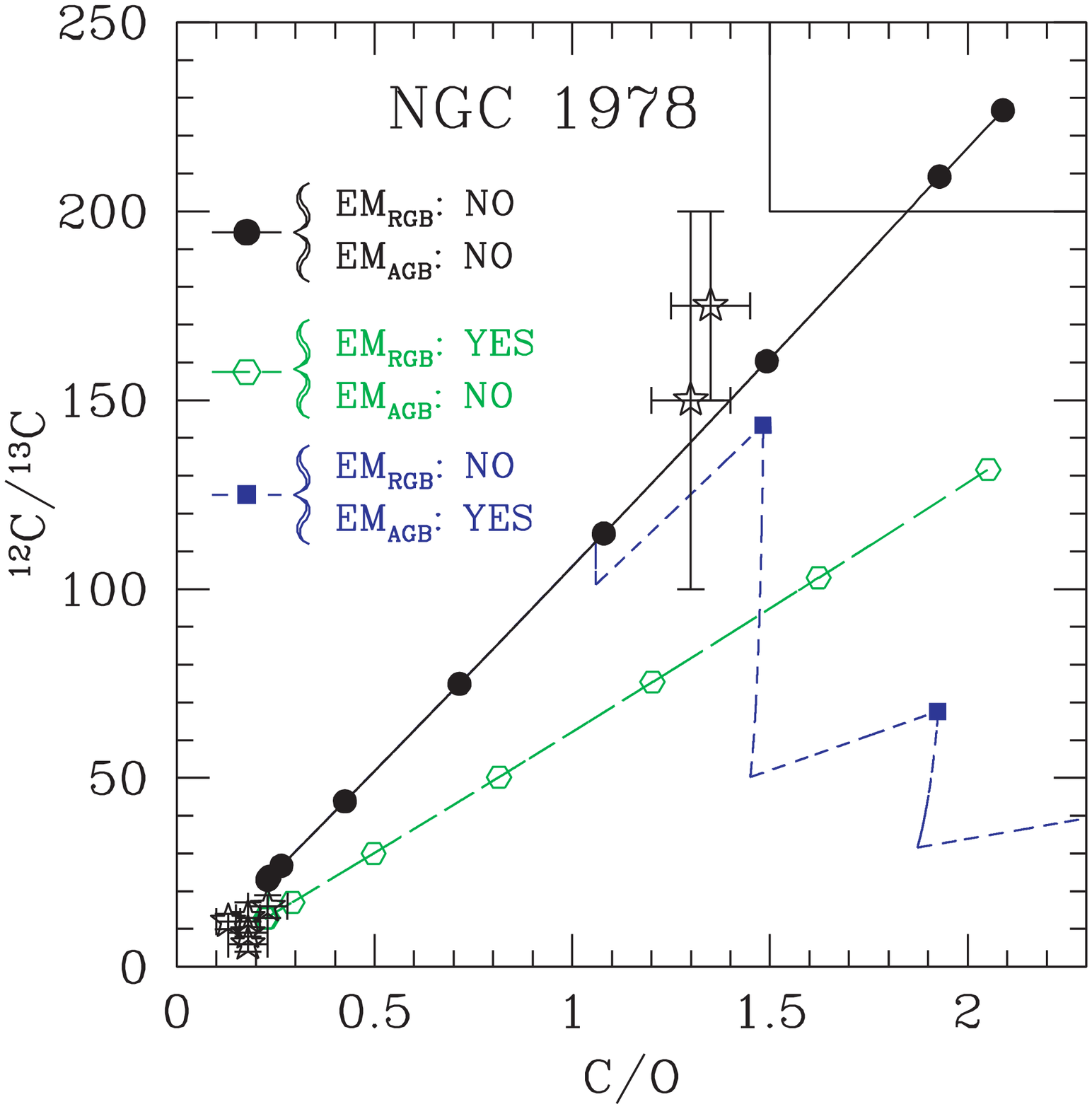}
\includegraphics[width=0.48\textwidth]{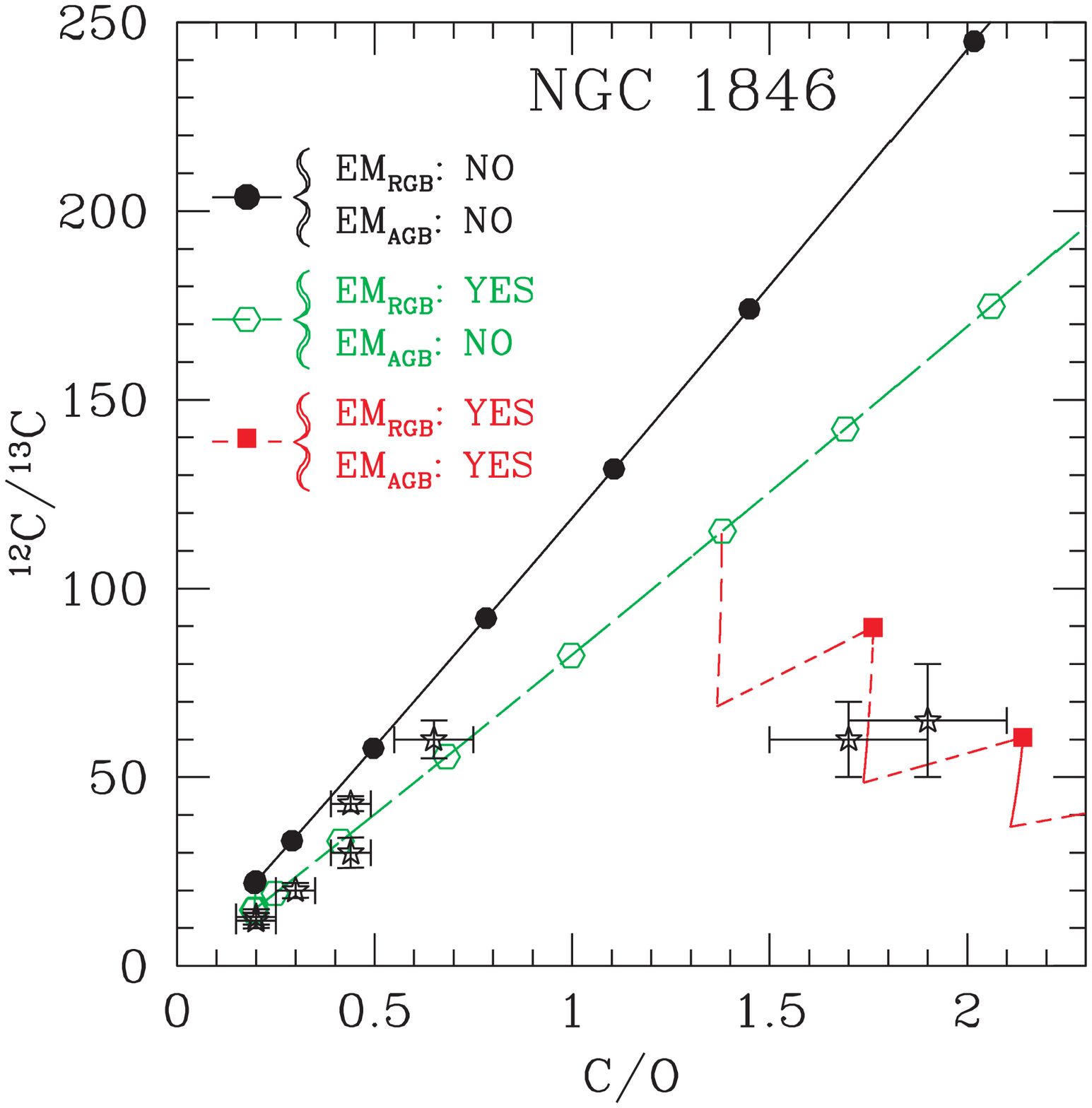}
   \caption{Comparison between observational data and theoretical
models. \element[][12]{C}/\element[][13]{C} isotopic ratios vs.
C/O ratios are reported. In the left panel, we report observational data of
the cluster NGC\,1978, accompanied by selected theoretical tracks. As a
comparison, in the right panel, we report the fitting curve for
another LMC cluster (NGC\,1846). The symbols along the tracks indicate the values attained during the interpulse phases (see also Fig.~\ref{fig:theo1}). The star symbols with error bars 
indicate our observations. See text for details.}
      \label{fig:theo2}
\end{figure*}

In the right panel of Fig.~\ref{fig:theo1} we report
the same models as shown in the left panel.
None of the theoretical tracks can simultaneously reproduce the abundance
ratios of the M and C stars. The high values of the C-star
carbon isotopic ratio can only be reproduced by the oxygen-enhanced
model with no RGB extra-mixing. All the other models predict too low
\element[][12]{C}/\element[][13]{C} ratios. The situation appears even more peculiar when compared
with the case of NGC\,1846, for which we found C stars with higher values of 
C/O, but with $\element[][12]{C}/\element[][13]{C}$ between $60$ and $70$, so that an AGB
extra-mixing (in addition to the RGB extra-mixing) was invoked to
reproduce the observations. The vast difference between the two
clusters is illustrated in Fig.~\ref{fig:theo2}. In the left panel,
data for NGC\,1978 are compared with three models computed under
different assumptions for the RGB and the AGB extra-mixing,
namely no extra-mixing (black-solid line), moderate RGB
extra-mixing only (green-long-dashed line), and moderate AGB
extra-mixing only (blue-dashed line). All the three models have $M=1.5\,M_\odot$ 
and $[\mathrm{O/Fe}]=0.2$. In the right panel, data for NGC\,1846
are compared with similar models. In this case, the mass is $1.8\,M_\odot$ and 
the red-dashed line refers to a model with both RGB
and AGB extra-mixing (see paper~I for more details on models
with AGB extra-mixing). In summary, the high carbon isotopic
ratio observed for the C stars of NGC\,1978 
($\element[][12]{C}/\element[][13]{C}>150$) rules out both
RGB and AGB extra-mixing, which are instead required to reproduce
the evolutionary sequence of M, S, and C stars in NGC\,1846.

\subsection{Alternative scenarios}\label{sect:alternativescenarios}
In the previous section, we did not identify a
theoretical scenario that satisfactorily reproduces the whole set of
stars belonging to NGC\,1978 (including both the O-rich and the
C-rich sub-sample). Moreover, the intermediate-age LMC cluster NGC\,1846, which we 
investigated in paper~I, presents a totally different abundance pattern,
which cannot be ascribed to the same theoretical scenario. The lack of
S stars is an additional peculiarity of NGC\,1978.
So far, our arguments have been based on the assumption that the stars in our sample
possess the same initial composition, mass, and age so that they form an evolutionary
sequence, in which the variations of the surface composition are due to 
intrinsic processes only.
In this section, we speculate about possible alternative scenarios, namely:
\begin{enumerate}
\item the existence of two stellar populations having different ages;
\item the existence of rejuvenated stars;
\item the activation of additional mixing at the base of the convective zones generated by a
thermal pulse.
\end{enumerate}

The existence of two populations proposed by
\citet{1999A&AS..135..103A} is an attractive possibility, because it
might explain the dichotomy between the M and the C stars. The M stars
would then belong to the older population. Their mass is sufficiently low for
the occurrence of the RGB extra-mixing, but too low for the
occurrence of the TDU during the AGB. On the contrary, the C stars
are younger (less than $1\,\mathrm{Gyr}$) and more massive ($M>2\,M_\odot$).
For these stars, the persistence of the $\mu$-barrier at the base
of the hydrogen-rich envelope prevented the RGB extra-mixing, whilst,
during the AGB, TDU accounts for the observed large C enhancement.
However, this scenario seems unlikely. First of all, a number
of other studies did not find significant metallicity, mass or age
differences. Second, a cluster merger creating two populations 
is not considered to be a valid option (see
Sect.~\ref{sect:observations}). \citet{1999A&AS..135..103A} argue 
that the north-western half could be more metal-rich and maybe 
slightly younger. The distribution of our targets in Fig.~\ref{fig:ngc1978} 
does not support this idea. The M stars are distributed equally around the
cluster centre. The carbon stars in our sample lie either in the
centre or in the south-eastern half of the cluster. According to
\citet{1980MNRAS.193...87L}, there is another carbon star (LE11)
located in the north-western half. Although the C stars appear to
line up along the cluster's major axis, there is no indication
that they concentrate in a certain region. A wide spread in
metallicity would have shown up in the fitting process, yet we
were able to obtain reasonable results for both oxygen-rich stars and
the hottest carbon-stars with a single value for the metallicity.

The scenario of a single age stellar population does not
exclude the existence of rejuvenated stars. Merging, coalescence,
or mass accretion processes are often invoked to explain the Blue
Stragglers observed in many GCs. Actually, a non-negligible number 
of stars lie above the turn-off and the sub-giant branch in the 
colour-magnitude diagram of NGC\,1978
\citep{2007AJ....133.2053M}, although we cannot exclude that those
are field stars or the result of unresolved pairs. 
If the product of one of the above mentioned
processes is a star with a mass larger than $2\,M_\odot$, or if the
rejuvenating process leaves the envelope composition enriched with
heavy elements, the efficiency of the RGB extra-mixing is
significantly reduced \citep{2007A&A...467L..15C}.
Such a scenario might explain the high carbon isotopic ratio of
the C stars, but the lack of S stars remains a mystery.\footnote{Note
that our sample contains almost all AGB stars in this cluster (see Fig.~\ref{fig:cmd}).}
Moreover, {even if we do not have a detailed statistics of the 
number of binary systems in NGC\,1978\footnote{Estimated binary fractions in other LMC and SMC clusters range from 10 \citep[NGC\,1783, ][]{2007AJ....134.1813M}
to $\leq30$ per cent \citep{2007A&A...466..165C}.}, it seems unlikely that two merger
events occur nearly simultaneously to produce the stars B and LE6,
for instance.

\begin{figure}[t]
\includegraphics[width=0.48\textwidth]{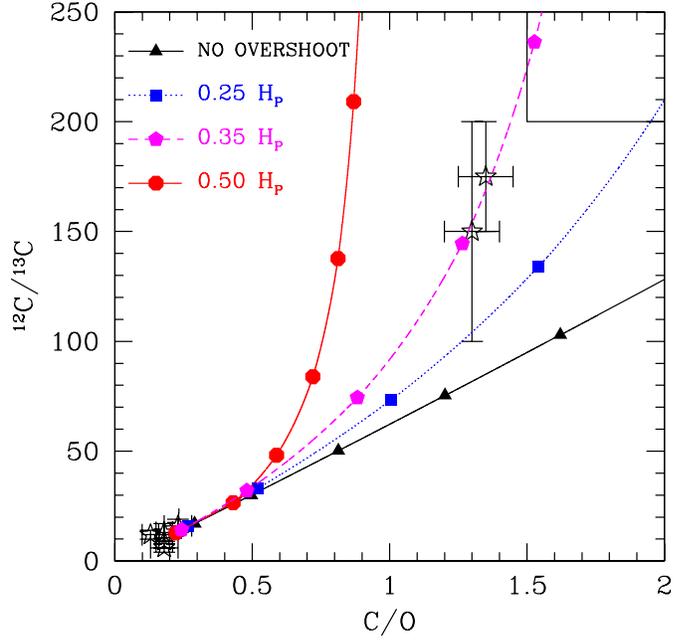}
\caption{\element[][12]{C}/\element[][13]{C} isotopic ratio vs.
C/O ratio for a series of models including an overshoot at the base of
convective zone generated by the TPs. The extension (in fraction of the pressure scale height)
of the overshoot zone is indicated in the legend.}
\label{fig:overshoot}
\end{figure}

The last alternative scenario we have investigated regards
the possible modification of the efficiency of the mixing
processes taking place during the AGB phase,
 which is responsible for the
increase of \element[][12]{C} and \element[][16]{O} in the envelope. A deeper TDU
would increase both the carbon isotopic ratio and the C/O ratio, thus leaving
the slope of the curves in the right panel of
Fig.~\ref{fig:theo1} unaltered. 
In this scenario, the possibility of producing S stars 
as a consequence of the TDU is reduced, since the transition to
the C-star stage is more rapid. We recall that
the TDU depth basically depends on the strength of the thermal
pulses: stronger TPs push outward the external layers more easily, therefore
inducing deeper dredge-up episodes. In turn, the strength of a
thermal pulse depends on many physical and chemical properties, such as
the core and the envelope mass, or the CNO content in the envelope
\citep{2003PASA...20..389S}. On the other hand, an increase of the
primary oxygen that is dredged up might limit the increase of the
C/O at the surface and steepen the \element[][12]{C}/\element[][13]{C}-C/O 
relation. The fraction of oxygen produced by the 
$\element[][12]{C}+\alpha$ reaction increases toward
the centre. Therefore, an additional mixing process, as due to a
ballistic overshoot or to non-standard mechanisms (such as
rotation or magnetic kick), that moves primary O (and C) from the deep
interior to the He-rich inter-shell, may indirectly explain the
high carbon isotopic ratio of the carbon stars in NGC\,1978. 
To shed light on this possibility, we computed some additional
models, in which we applied an artificial overshoot at the base of
the convective zone generated by the TPs. The initial mass adopted 
in the models is $1.5\,M_\odot$, and the initial composition is 
oxygen-enhanced ($+0.2$~dex). A moderate RGB extra-mixing 
($T^\mathrm{max}_\mathrm{RGB}=2.1\times10^7$\,K, corresponding to 
the case of the green long-dashed line in Fig.~\ref{fig:theo1}) has
been included. We parametrise the additional mixing by limiting the 
extension of the overshoot layer to a fraction of the pressure scale 
height of the most internal convective mesh point. 
In Fig.~\ref{fig:overshoot}, we report the resulting 
\element[][12]{C}/\element[][13]{C} isotopic ratios as a function of C/O.

Even if these models apparently provide a simultaneous
reproduction of the composition of both M and C stars, we have to
stress the theoretical and observational evidence against
such a process. At the He-flash peak, the inner border of the
convective zone already reaches the layer where the temperature attains its
maximum value. Below this point, due to the neutrino energy loss, the
temperature (and the entropy) decreases, so that the deceleration
due to the buoyancy is quite strong. \citet{2006ApJ...642.1057H} showed that
even if the pressure and the entropy barrier at the base of the 
convective shell generated by a TP is stiff, convective plumes can 
penetrate into the underlying radiative layers due to
{\it g}-mode oscillations. These structures reach down to about 
$2\times10^8$\,cm below the formal convective border (see their Fig. 24), 
and develop mean vertical velocities that are about $1000$ times lower 
than the average convective velocities. 
The extension of this penetration has the same order of
magnitude as the overshoot zone in our models, namely
$1.0\times10^7$\,cm, $1.2\times10^8$\,cm, and
$1.5\times10^8$\,cm, for $0.25$ $H_P$, $0.35$ $H_P$, and $0.5$ $H_P$,
respectively (these are mean values). However, if in our 1D models the whole 
mass of the spherical overshoot region is
efficiently mixed, the amount of mass contained in the
penetrating plumes of the 3D model is certainly lower. In
addition, the result of the 3D model may be affected by the 
assumed boundary conditions. As was stressed by \citet{2006ApJ...642.1057H}, their
models ignore $\mu$-gradients, which increase the
stability and reduce mixing across the formal convective boundaries. 
Moreover, they found that the amplitude of the gravity waves depends both on
the resolution adopted in the simulation and on the choice of the heating rate.

On the other hand, the bulk of the C stars in the Milky Way as well
as those in NGC\,1846 have \element[][12]{C}/\element[][13]{C} ratios in the range from $40$ to $70$,
and only in a few cases up to $100$ \citep{1986ApJS...62..373L}. This occurrence leads us to the conclusion
that such an overshoot is supposedly an uncommon process.
Note that the occurrence of an overshoot below the convective zone
generated by a TP causes considerable changes of the physical
conditions in the He-rich inter-shell. In particular, higher 
temperatures develop at the pulse peak and, therefore, the
$\element[][22]{Ne}(\alpha,\mathrm{n})\element[][25]{Mg}$
reaction rate becomes an important neutron source, even in low-mass AGB stars.
In such a case, the resulting {\it s}-process nucleosynthesis would 
be characterised by an overproduction of neutron-rich isotopes. For 
instance, the expected isotopic compositions of strontium, zirconium, 
molybdenum, and barium would substantially differ from those
measured in pre-solar SiC grains, which were produced in the
cool circumstellar envelope of a past generation of C stars 
\citep[see][]{2003ApJ...593..486L}. {Additionally, calculated element 
abundances ratios such as Rb/Sr would not match 
observations obtained from carbon stars
\citep[see][]{2001ApJ...559.1117A}.}
For these reasons, if the
overshoot from the base of the convective zone
generated by TPs is non-negligible, it should be a rare event.

In summary, a clear and coherent picture of the AGB stars
in the cluster NGC\,1978 within the more general context of low-mass 
AGB stellar evolution and nucleosynthesis cannot be easily
achieved. Once again, we stress the uniqueness of the C star
sample of NGC\,1978 with respect to other LMC clusters.

\section{Conclusions}\label{sect:conclusions}
In this paper, we presented a sample of AGB stars
belonging to the LMC cluster NGC\,1978. Where it was possible, we
derived the C/O and the \element[][12]{C}/\element[][13]{C}
ratios. The spectroscopic data reveal the presence of two sub-samples,
one where $\mbox{C/O}<1$ in the stellar atmosphere (oxygen-rich or M-type stars), 
and the other with $\mbox{C/O}>1$ (carbon-rich or C-type stars). The oxygen-rich 
stars present low values for the C/O ratio
(mean value $0.18$) and the \element[][12]{C}/\element[][13]{C} ratio (mean value $11$).
The spread in the C/O ratio is restricted, all values are
consistent with the mean value within the error bars. We observe a wider
spread in the \element[][12]{C}/\element[][13]{C} ratios.
We conclude from the spectroscopic data in comparison to our evolutionary models 
that the TDU mechanism is not (yet) working in these stars.
The carbon-rich stars of our sample instead show C/O
values and \element[][12]{C}/\element[][13]{C} ratios consistent with the occurrence
of TDU. A lack of S stars, with C/O ratios
between those of the O-rich stars and the C-rich stars, has been highlighted.

We did not find a theoretical scheme that is able to satisfactorily 
reproduce the chemical abundance pattern in NGC\,1978. By claiming 
the existence of a non-standard mixing mechanism, we postulated some 
possible solutions. However, we are aware that too many {\em ad hoc} 
assumptions make our analysis objectionable, taking into account that 
some of them are strongly limited by observational constraints. In 
particular, a more consistent picture of all the additional mixing 
mechanisms, which are active during the RGB and the AGB phase, is required.
We also discussed the scenario of multiple stellar populations and
rejuvenated stars, and highlighted the differences to the 
abundance pattern in the cluster NGC\,1846.

Narrower constraints for stellar evolutionary models could be derived 
from atmospheric models including a set of accurate molecular line data,
especially in the case of carbon stars. Additionally, the more evolved
a star is (thus having a higher luminosity and C/O ratio), the more dynamic
effects influence the spectral appearance. From the observational
point of view, improvements could be achieved by means of the choice of
the observed wavelength regions, some of which are less crowded with molecular
lines. Moreover a comparison between low-resolution data and high-resolution data
should be explored further \citep[cf.][]{2006MmSAI..77..955W}. If low-resolution 
data deliver robust C/O ratios (and stellar parameters), the isotopic ratios 
could be derived in the high-resolution spectra more accurately.

In conclusion, we affirm that there is much room for improvement.
As a first step, we are currently acquiring new data to increase our statistics 
by studying more AGB stars in other LMC globular clusters. In this way, we could
shed light on the puzzling abundance pattern that we found in NGC\,1978.

\begin{acknowledgements}
MTL and TL acknowledge funding by the Austrian Science Fund FWF
(projects {P-18171} and {P-20046}). MTL has been supported by the
Austrian Academy of Sciences (DOC programme). OS and SC have been supported by the
MIUR Italian Grant Program PRIN 2006. BA acknowledges funding by 
the Austrian Science Fund FWF (project {P-19503}). 
Based on observations obtained at the Gemini Observatory, which is operated by the
Association of Universities for Research in Astronomy, Inc., under a cooperative agreement
with the NSF on behalf of the Gemini partnership: the National Science Foundation (United
States), the Science and Technology Facilities Council (United Kingdom), the
National Research Council (Canada), CONICYT (Chile), the Australian Research Council
(Australia), Minist\'{e}rio da Ci\^{e}ncia e Tecnologia (Brazil) 
and Ministerio de Ciencia, Tecnolog\'{i}a e Innovaci\'{o}n Productiva  (Argentina).
The observations were obtained with the Phoenix
infrared spectrograph, which was developed and is operated by the
National Optical Astronomy Observatory. The spectra were obtained
as part of the programs GS-2006B-C-6 and GS-2008A-Q-65. This
publication makes use of data products from the Two Micron All Sky
Survey, which is a joint project of the University of
Massachusetts and the Infrared Processing and Analysis
Center/California Institute of Technology, funded by the National
Aeronautics and Space Administration and the National Science
Foundation.
\end{acknowledgements}

\bibliographystyle{aa}
\bibliography{lederer-ngc1978}

\begin{appendix}
\section{Changes to the CO line list}\label{sect:colist}
To match the positions of the CO lines in the recorded K-band spectra, we had to modify some transition frequencies as given in the list of \citet{1994ApJS...95..535G}\footnote{Available at the CDS: {\tt http://vizier.cfa.harvard.edu/viz-bin/VizieR?-source=J/ApJS/95/535}.}. The wavenumbers were taken from the Infrared Atlas of the Arcturus Spectrum \citep{1995iaas.book.....H}. The changes are documented in Table~\ref{table:colines}. The given quantities are the old and new transition frequency ($\sigma$), the lower state term energy $E_\mathrm{l}$, the logarithm of the $gf$-value ($\log gf$), the lower and upper state vibrational quantum numbers ($v_\mathrm{l}$ and $v_\mathrm{u}$), the lower state angular momentum number ($J_\mathrm{l}$), and the isotope identification.

At a later point in the analysis, we compared the corrections of the line positions to the CO data given by \citet{1996A&AS..117..557C}. The wavenumbers from the Arcturus atlas are in good agreement with their transition frequencies. The differences are less than $0.04$ cm$^{-1}$ for the 1-3 vibration transitions and less than $0.01$ cm$^{-1}$ for the other lines. However, we did not use the \citet{1996A&AS..117..557C} data, because the line strengths of the \element[][12]{CO} 3-0 band head in the H band from this list are severely underestimated and therefore unusable for the analysis. 

\begin{table*}
\caption{Modified line positions of CO lines in the K band}
\label{table:colines}
\centering
\begin{tabular}{llllllll}
\hline\hline
$\sigma_\mathrm{old}$ [cm$^{-1}$] & $\sigma_\mathrm{new}$ [cm$^{-1}$] & $E_\mathrm{l}$ [cm$^{-1}$] & $\log gf$ & $v_\mathrm{l}$ & $v_\mathrm{u}$ & $J_\mathrm{l}$ & Isotope \\
\hline
4219.5583 & 4225.3630 & 19009.6550 & 4.865E-05 & 1 & 3 & 95 & \element[][12]{C}\element[][16]{O} \\
4226.9718 & 4227.7190 & 14547.6172 & 6.569E-05 & 2 & 4 & 74 & \element[][12]{C}\element[][16]{O} \\
4223.6261 & 4228.9450 & 18668.6535 & 4.782E-05 & 1 & 3 & 94 & \element[][12]{C}\element[][16]{O} \\
4228.9365 & 4229.6010 & 14278.1804 & 6.439E-05 & 2 & 4 & 73 & \element[][12]{C}\element[][16]{O} \\
4230.8124 & 4231.4030 & 14012.1198 & 6.310E-05 & 2 & 4 & 72 & \element[][12]{C}\element[][16]{O} \\
4227.5739 & 4232.4430 & 18330.8030 & 4.701E-05 & 1 & 3 & 93 & \element[][12]{C}\element[][16]{O} \\
4232.1241 & 4233.3140 & 11962.5127 & 1.167E-05 & 0 & 2 & 81 & \element[][13]{C}\element[][16]{O} \\
4234.3009 & 4234.7650 & 13490.1716 & 6.057E-05 & 2 & 4 & 70 & \element[][12]{C}\element[][16]{O} \\
4234.3757 & 4235.4470 & 11676.5472 & 1.144E-05 & 0 & 2 & 80 & \element[][13]{C}\element[][16]{O} \\
4235.9151 & 4236.3250 & 13234.3055 & 5.932E-05 & 2 & 4 & 69 & \element[][12]{C}\element[][16]{O} \\
4237.4434 & 4237.8040 & 12981.8588 & 5.809E-05 & 2 & 4 & 68 & \element[][12]{C}\element[][16]{O} \\
4238.8865 & 4239.2040 & 12732.8418 & 5.687E-05 & 2 & 4 & 67 & \element[][12]{C}\element[][16]{O} \\
4240.2450 & 4240.5230 & 12487.2648 & 5.566E-05 & 2 & 4 & 66 & \element[][12]{C}\element[][16]{O} \\
4241.5194 & 4241.7630 & 12245.1376 & 5.447E-05 & 2 & 4 & 65 & \element[][12]{C}\element[][16]{O} \\
\hline
\end{tabular}
\end{table*}

\end{appendix}

\end{document}